\documentclass[%
superscriptaddress,
showpacs,preprintnumbers,
bibnotes,
amsmath,amssymb,
aps,
pra,
twocolumn
]{revtex4-1}

\usepackage{graphicx}% Include figure files
\usepackage{dcolumn}% Align table columns on decimal point
\usepackage{bm}% bold math
\usepackage{braket}
\usepackage[utf8]{inputenc}
\usepackage{xcolor}

\begin{document}

%\preprint{APS/123-QED}

%\title{From sudden quenches and thermal annealing to adiabatic switching in multimode cavity induced atomic selfordering}% Force line breaks with \\
%\thanks{A footnote to the article title}%

\title{Quenches across the self-organization transition in multimode cavities}

\author{Tim Keller}
\affiliation{Theoretische Physik, Universit\"at des Saarlandes, D-66123 Saarbr\"ucken, Germany}

\author{Valentin Torggler}
\affiliation{Institut f\"ur Theoretische Physik, Universit\"at Innsbruck, A-6020 Innsbruck, Austria}

\author{Simon B. J\"ager}
\affiliation{Theoretische Physik, Universit\"at des Saarlandes, D-66123 Saarbr\"ucken, Germany}

\author{Stefan Sch\"utz}
\affiliation{Theoretische Physik, Universit\"at des Saarlandes, D-66123 Saarbr\"ucken, Germany}
\affiliation{icFRC, IPCMS (UMR 7504) and ISIS (UMR 7006), University of Strasbourg and CNRS, 67000 Strasbourg, France}

\author{Helmut Ritsch}
\affiliation{Institut f\"ur Theoretische Physik, Universit\"at Innsbruck, A-6020 Innsbruck, Austria}

\author{Giovanna Morigi}
\affiliation{Theoretische Physik, Universit\"at des Saarlandes, D-66123 Saarbr\"ucken, Germany}

%\collaboration{Collaboration}%\noaffiliation

\date{\today}% It is always \today, today,
             %  but any date may be explicitly specified

\begin{abstract}
A cold dilute atomic gas in an optical resonator can be radiatively cooled by coherent scattering processes when the driving laser frequency is tuned close but below the cavity resonance. When sufficiently illuminated, moreover, the atoms' steady state undergoes a phase transition from homogeneous density to crystalline order. We characterize the dynamics of this self-ordering process in the semi-classical regime when distinct cavity modes with commensurate wavelengths are quasi-resonantly driven by laser fields via scattering by the atoms. The lasers are simultaneously applied and uniformly illuminate the atoms, their frequencies are chosen so that the atoms are cooled by the radiative processes, their intensity is either suddenly switched or slowly ramped across the self-ordering transition. Numerical simulations for different ramp protocols predict that the system exhibits long-lived metastable states, whose occurrence strongly depends on initial temperature, ramp speed, and number of atoms. 
\end{abstract}

%\pacs{Valid PACS appear here}% PACS, the Physics and Astronomy
                             % Classification Scheme.
%\keywords{Suggested keywords}%Use showkeys class option if keyword
                              %display desired
\maketitle

\section{Introduction}

Laser light creates an attractive optical potential for cold atoms when far detuned below an optical transition. Such potential can be significantly enhanced if the light is confined by an optical resonator \cite{Horak:1997,Domokos:2003,Black:2003,Baumann:2010}. In addition, if the laser illuminates the atoms, trapping is induced by a dynamical optical potential emerging from the interference between the scattered light and the laser, which tends to order the particles at the maxima of the intensity \cite{Asboth:2005,Baumann:2010}. The interference contrast, and thus the trapping depends on the relative positions of the scattering atoms. Therefore, this phenomenon can be also understood in terms of an effective long-range force, which is mediated by the collectively scattered photons \cite{Asboth:2005,Schuetz:2014,Schuetz:2015,ODell:2003,Muenstermann:2000}. This force has also a dissipative component, which is due to the dissipative nature of the resonator and which cools the atoms when the pump is tuned below the cavity resonance \cite{Vuletic:2000,Black:2003}. Theoretical studies with single-mode resonators predicted that this dissipation can establish long-range correlations and support the onset of metastable ordered structures \cite{Schuetz:2016,Jaeger:2016}.

\begin{figure}
\centering
\includegraphics[width=0.48\textwidth]{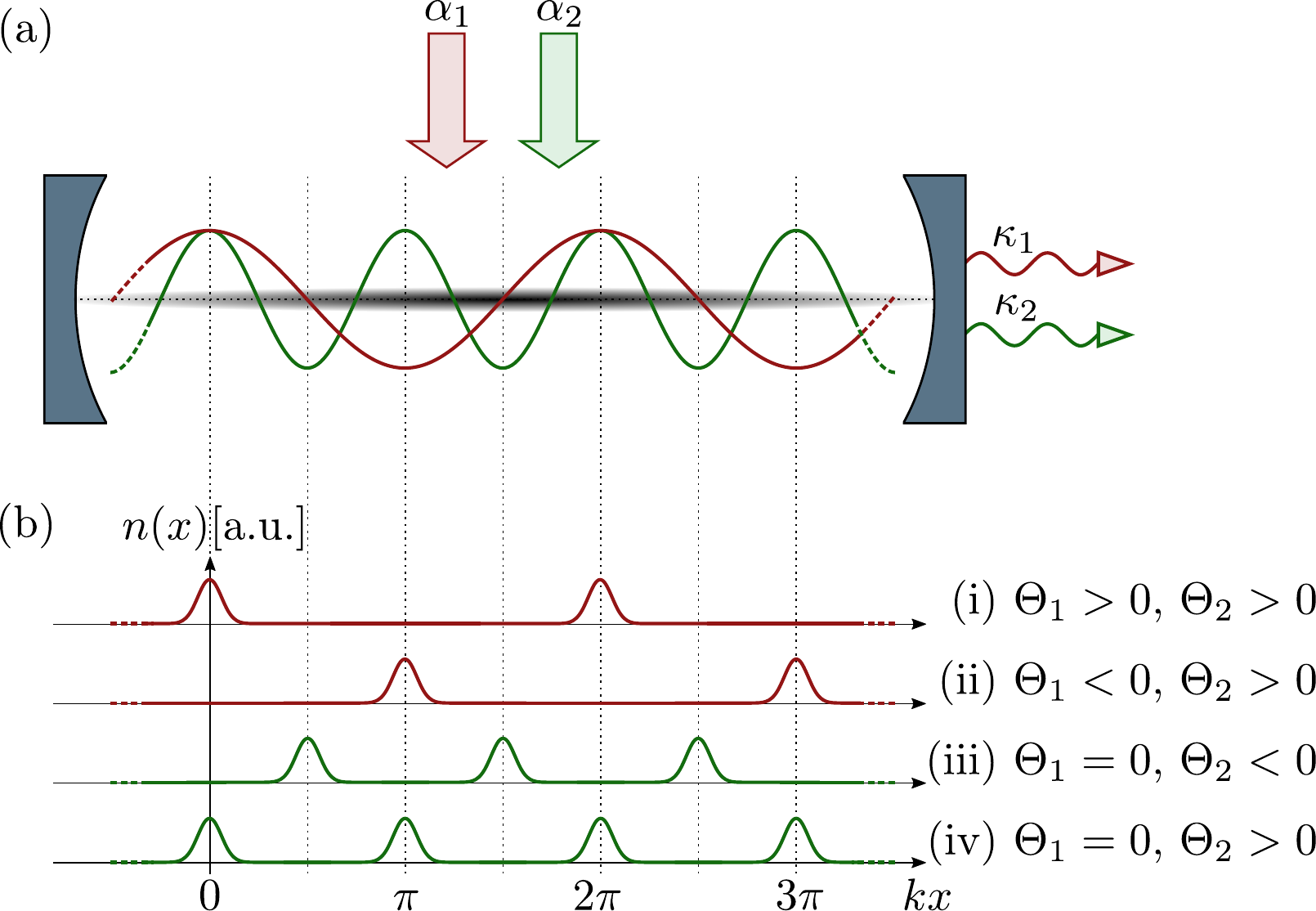}
\caption{\textbf{(a)} Cold atoms are confined within an optical cavity and move along the cavity axis ($x$ axis). They coherently scatter photons from transverse lasers with rescaled amplitudes $\alpha_1$ and $\alpha_2$ into the (correspondingly) resonant cavity modes with spatial mode functions $\cos(kx)$ (red) and $\cos(2kx)$ (green) and loss rates $\kappa_1$ and $\kappa_2$, respectively. 
\textbf{(b)} Sketch of the atomic density distribution $n(x)$ along the cavity axis (in units of $1/k$) for the four possible stationary self-organized orders. On the right we report the corresponding values of the quantities $\Theta_1$ and $\Theta_2$, signaling Bragg order in the mode 1 and 2, respectively. See text and Ref. \cite{Keller:2017} for details.}
\label{Fig:1}
\end{figure}

In a multimode cavity and for several illumination frequencies, competing ordering processes are present and lead to a richer phase dynamics. In a two-mode cavity, like the one depicted in Fig.\ \ref{Fig:1}(a), the transition to self-organization can be a phase-transition of first or second order depending on the laser intensities and on their relative strength \cite{Keller:2017}. The corresponding self-ordered phases can exhibit superradiant scattering either in one or in both cavity modes, as illustrated in Fig.\ \ref{Fig:1}(b), while the asymptotic distribution of the atoms can be thermal provided that the lasers' frequencies are suitably chosen \cite{Keller:2017}. In our example the particles can order in a lattice at a given length scale $\lambda$ and/or on a lattice with half the period $\lambda/2$. For these settings we numerically analyze the semi-classical dynamics following sudden quenches or slow ramps of the laser intensities across the thresholds separating the homogeneous from one of the self-organized phases. We describe the evolution by stochastic differential equations, which correspond to the Fokker-Planck equation derived in Ref.\ \cite{Schuetz:2013} for a similar system. We find that even at very long times the atoms' spatial distribution strongly depends on the initial temperature, ramp speeds, and on the quench protocol, such that the system gets trapped in long-lived metastable states. In particular, for quenches starting with ensembles at low temperatures, the buildup of long-range order requires longer times when compared to higher initial temperatures. 

Our work is organized as follows: In Section \ref{sec:system} we introduce the system and the semi-classical equations describing the dynamics. The atoms' stationary properties are then summarized in a phase diagram, which was derived in Ref.\ \cite{Keller:2017}. In Sec.\ \ref{sec:dynamics} we numerically study the real time dynamics when the parameters are varied within the phase diagram according to different quench protocols. In Sec.\ \ref{sec:mpemba} we analyze the crystallization dynamics starting from spatially homogeneous distributions with different momentum widths. In Sec.\ \ref{sec:comparison} we compare the predictions of the stochastic differential equations we employ with an extended approach including the dynamical evolution of the field modes introduced in Refs. \cite{Domokos:2001,Torggler:2014}. The conclusions are drawn and future perspectives are discussed in Sec.\ \ref{sec:conclusions}.

\section{Semiclassical dynamics}
\label{sec:system}

The system we consider consists of a gas of $N$ cold atoms with mass $m$, which are trapped inside a high-finesse optical resonator and coherently scatter laser light into the cavity modes. The atomic motion is confined along the cavity axis (here the $x$ axis) by a tight external dipole trap \cite{Brennecke:2007,Landig:2016} and is here described in the semi-classical limit. 

The geometry of the setup is illustrated in Fig.\ \ref{Fig:1}. Lasers with (rescaled) intensities $\alpha_n$  propagate in a direction orthogonal to the cavity axis and are quasi resonant with the standing wave cavity modes $\cos(nkx)$ with frequency $\omega_{c,n}$ and wave number $nk$ ($n=1,2$) \cite{footnote:1,Leonard:2017}. The lasers have frequency $\omega_{p,n}$ and linear polarization which is parallel to the one of the corresponding cavity mode. Each pair of laser and cavity mode couples to an atomic dipolar transition at frequency $\omega_{a,n}$, where $\Omega_{p,n}$ and $g_n$ are the laser and the vacuum Rabi frequency, respectively. Spontaneous scattering processes are suppressed when the absolute value of the detuning $\Delta_{a,n}=\omega_{p,n}-\omega_{a,n}$ exceeds the coupling strengths and the detuning $\Delta_n=\omega_{p,n}-\omega_{c,n}$ between laser and cavity mode by orders of magnitude: $|\Delta_{a,n}|\gg \Omega_{pn,},g_n, |\Delta_n|$. The relevant dissipative process are given by cavity decay, and we denote by $\kappa_n$ the loss rate of cavity mode $n=1,2$.

In the so-called bad cavity limit, assuming that the cavity field loss rates are faster than the rate of the dynamics of the atomic motion, one can eliminate the cavity field variables from the equations of motion of the atoms by means of a coarse graining in time. This gives rise to an effective model, where the atoms experience a long-range interaction mediated by the cavity photons, while retardation effects and fluctuations of the cavity field are responsible for friction forces and diffusion. In the semi-classical limit one can derive a Fokker-Planck equation for the atoms' position and momentum distribution, assuming that the single-atom momentum distribution has a width $\Delta p$ which, at all instants of time, is orders of magnitude larger than the photon recoil $\hbar k$: $\Delta p\gg \hbar k$ \cite{Dalibard:1985,Schuetz:2013,Keller:2017}.
The corresponding stochastic differential equations read
\begin{subequations}
\begin{align}
dx_j &=\frac{p_j}{m}dt\,,\\
dp_j &= (F_{j,\mathrm{ad}}+F_{j,\mathrm{ret}})dt + dW_j^{(1)} + dW_j^{(2)} \,,\label{forceel}
\end{align}
\label{com}
\end{subequations}
where
\begin{align}
	F_{j,\mathrm{ad}}=&-\sum_{n=1,2}2\hbar nk \frac{\alpha_n}{\hbar \beta_n}  \sin(nk x_j)\Theta_n\label{Forcead}\,,\\
	F_{j,\mathrm{ret}}=&-\sum_{n=1,2}\frac{\hbar (nk)^2}{m} \alpha_n \frac{\kappa_n}{-\Delta_n}\sin(nk x_j)\frac{1}{N}\sum_{l=1}^Np_l\sin(nkx_l)\,, \label{Forceret}
\end{align}	
and
\begin{align}
\alpha_n=&\frac{4NS_n^2\Delta_n^2}{(\Delta_n^2+\kappa_n^2)^2}\label{eq:nthreshold}\,,\\
\beta_n=&\frac{-4\Delta_n}{\hbar (\Delta_n^2+\kappa_n^2)}\label{eq:nbeta}\,.
\end{align}
Here, $S_n=g_n\Omega_n/\Delta_{an}$ is the amplitude of coherent scattering by a single atom and has the dimension of a frequency, while $dW_j^{(1)}$ and $dW_j^{(2)}$ in Eq.\ \eqref{forceel} describe Wiener processes, which fulfill $\langle dW_i^{(n)}\rangle =0$ and $\langle dW_i^{(n)} dW_j^{(m)}\rangle = 2 D_{ij}^n\delta_{nm} dt$ ($n,m=1,2$ and $i,j=1,...,N$). Here,
\begin{equation}
D_{ij}^n=   (\hbar nk)^2\frac{\alpha_n}{\hbar\beta_n}\frac{\kappa_n}{-\Delta_n}\sin(nkx_i)\sin(nkx_j)\,.
\label{eq:diffusion}
\end{equation}
Finally, the parameter 
\begin{equation}
\Theta_n=\frac{1}{N}\sum_{i=1}^N\cos(nkx_i)
\end{equation}
quantifies Bragg ordering of the atoms in the cavity mode with wave number $nk$. In particular, $|\Theta_n|=1$ when the atoms are localized either at the maxima or at the minima of $\cos(nkx)$, which is the configuration which maximizes the intracavity field intensity. We identify $\Theta_n$ with the order parameter for self-organization in the corresponding cavity mode \cite{Keller:2017}. Below, we will denote by ''long-wavelength order'' a configuration with non-vanishing value of $\Theta_1$, corresponding to a Bragg grating with period $\lambda=2\pi/k$. Similarly, ''short-wavelength order'' refers to a configuration with $\Theta_2\neq0$, corresponding to a Bragg grating with $\lambda/2$. Note that here and in the rest of the paper we discard the dynamical Stark shift of the cavity frequency assuming that this is much smaller than the cavity mode linewidth $Ng_{n}^2/|\Delta_{a,n}|\ll\kappa_n$. For details we refer to Ref. \cite{Keller:2017}.

\subsection{Stationary states}
\label{sec:stationaryproperties}

An analysis of the Fokker-Planck equation at the basis of Eq.\ \eqref{com} allows to identify the conditions for the existence of a stationary state. The latter exists provided that $\Delta_n<0$ and $\beta_1=\beta_2\equiv \beta$, see Eq.\ \eqref{eq:nbeta}. In this case the atoms' distribution at steady state reads \cite{Keller:2017}
\begin{align}
\label{f:st}
	f_{\mathrm{st}}(x_1,p_1,\ldots,x_N,p_N)=\frac{\exp(-\beta H_{\mathrm{eff}})}{\mathcal Z(\beta)}
\end{align}
where $H_{\mathrm{eff}}$ is the effective Hamiltonian derived after eliminating the cavity field variables, 
\begin{align}
	H_{\mathrm{eff}}=\sum_{j=1}^N\frac{p_j^2}{2m}-\sum_{n=1,2}N\frac{\alpha_n}{\beta_n}\Theta_n^2\label{eq:hamiltonian}\,,
\end{align}
while $\mathcal Z(\beta)$ denotes the  partition function:
\begin{align}
\mathcal Z(\beta)=\frac{1}{\Delta^N}\int_{-\infty}^{\infty} dp_1\,...\,\int_{-\infty}^{\infty} dp_N\int_{0}^{\lambda} dx_1\,...\,\int_{0}^{\lambda} dx_N\,e^{-\beta H_{\mathrm{eff}}}\,,
\end{align}
with $\lambda=2\pi/k$ and with $\Delta=2\pi\hbar$ the single particle unit phase space volume. In the following we will assume that the cavity decay rates are equal, 
\begin{align}
\kappa_1=\kappa_2=:\kappa\label{kappaequal}\,,
\end{align}
so that the condition for the existence of the stationary state in Eq.\ \eqref{f:st} becomes
\begin{align}
\Delta_1=\Delta_2=:\Delta_c<0\label{Deltacequal}\,.
\end{align} 
The phase diagram of the system can be determined by using that the steady state, Eq.\ \eqref{f:st}, has the form of a thermal state. On the basis of this observation we introduce the temperature $T$ of the stationary state, which is defined as
\begin{align}
k_{B}T=\beta^{-1}=\frac{\hbar (\Delta_c^2+\kappa^2)}{-4 \Delta_c}\,, \label{T}
\end{align}
with the Boltzman constant $k_B$. The steady-state temperature $T$ has the same functional dependence on $\Delta_c$ and $\kappa$ as for a single-mode cavity \cite{Schuetz:2013,Schuetz:2015}. We can further define the free energy per particle $\mathcal{F}$ using the formal equivalence with the canonical ensemble of equilibrium statistical mechanics \cite{Schuetz:2015}:
\begin{align}
\mathcal{F}=-\frac{1}{N\beta}\ln(\mathcal Z(\beta))\label{freeenergy}\,.
\end{align}
Following the procedure detailed in Refs.\ \cite{Schuetz:2015,Keller:2017,Jaeger:2016} we determine the global minima of $\mathcal{F}$ in an appropriately defined thermodynamic limit, which consists in keeping $\alpha_n$ constant for $N\rightarrow\infty$. The global minima are the resulting stationary phases. The corresponding order parameters $\Theta_1$ and $\Theta_2$, in particular, are determined by $\alpha_1$ and $\alpha_2$. When  the fields are sufficiently weak, then $\Theta_1=\Theta_2=0$ the density is homogeneous and there is no structural order. We denote this phase by paramagnetic, borrowing the notation of the generalized Hamiltonian mean-field model (GHMF) \cite{Campa:2009,Teles:2012,Pikovsky:2014} to which this model can be mapped. The possible ordered phases at steady state are illustrated in Fig.\ \ref{Fig:1}(b) and take one of four set of values. In particular, the ferromagnetic phase is characterized by (i) $\Theta_1>0$, $\Theta_2>0$ and (ii) $\Theta_1<0$, $\Theta_2>0$, exhibiting Bragg order in both cavity modes. The nematic phases (iii) and (iv), instead, are characterized by no order in the long-wavelength mode, $\Theta_1=0$ while $\Theta_2$ can be either negative or positive.

\begin{figure}
\centering
\flushleft(a)\\
\center
\includegraphics[width=0.44\textwidth]{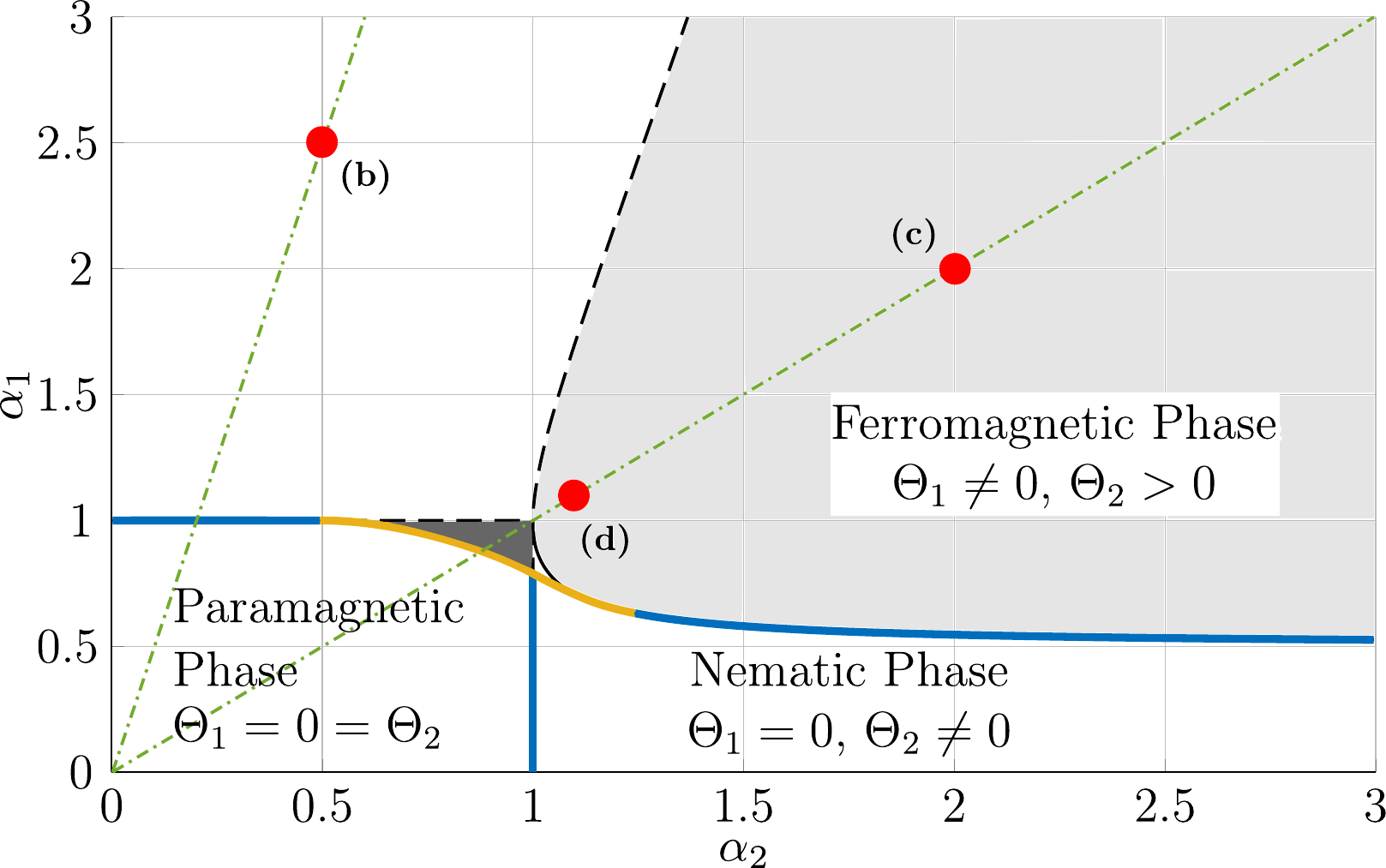}\\
\flushleft(b) \hspace{3.8cm}(c)\\
\includegraphics[width=0.23\textwidth]{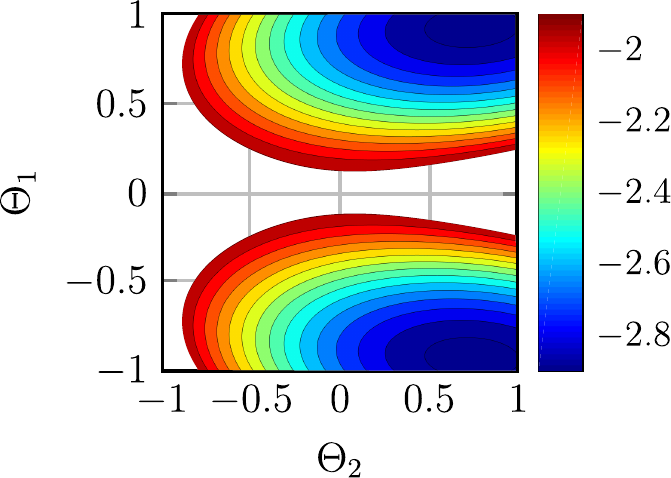}
\includegraphics[width=0.23\textwidth]{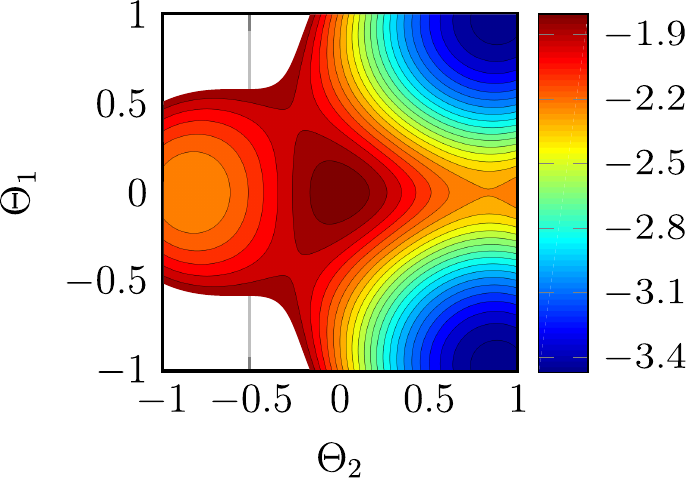}
\caption{(a) Phase diagram of the stationary phases, corresponding to the global minima of Eq. \eqref{freeenergy}, in the plane $\alpha_1-\alpha_2$. Blue (yellow) lines mark second (first) order phase transitions. The light gray area (dark gray area) within the ferromagnetic phase indicate the parameter region where the nematic (paramagnetic) phase are local minima of the free energy. The red circles labeled by (b) and (c) indicate the parameters to which the contour plots of free energy in subplots (b) and (c) are shown in the $\Theta_1-\Theta_2$ landscape. The free energy in subplot (b) exhibits two global minima at $\Theta_1=\pm 0.92$ and  $\Theta_2=0.73$; in (c) the two global minima of the ferromagnetic phase are at $\Theta_1=\pm 0.97$ and $\Theta_2=0.88$, the local minimum in the nematic phase is at $\Theta_1=0$ and $\Theta_2=-0.83$ (the contour of $\mathcal{F}$ is reported below a convenient threshold). The green dash-dotted lines in subplot (a) illustrate the paths of the quench protocols discussed in Sec. \ref{sec:dynamics}. Circle (d) indicates the parameters of the quench discussed in Sec. \ref{subsec:slowramp}.  
 }
\label{fig:phase_diagram}
\end{figure}

The resulting phase diagram in the $\alpha_1-\alpha_2$ plane is shown in Fig.\ \ref{fig:phase_diagram} and reproduces the one of Ref.\ \cite{Keller:2017}. The phases are separated by either first- or second-order transitions, which have been determined using Ehrenfest's criterion \cite{Pikovsky:2014}. The shaded areas show stability regions in which the free energy has a local minimum that corresponds to the paramagnetic (dark gray region) and nematic (light gray region) phase. Examples of the free-energy landscape in the $\Theta_1-\Theta_2$ plane are shown in subplots (b) and (c). Subplot (b) corresponds to the parameters of the red bullet labeled by (b) in subplot (a): Here, the free energy exhibits two symmetric global minima which correspond to the ferromagnetic phase. In subplot (c), corresponding to the parameters of the red bullet labeled by (c), there is an additional local minimum corresponding to a nematic phase. In the latter there is only ordering in the short-wavelength lattice, while $\Theta_1=0$. We denote this region by {\it bistable} referring to the existence of a second, metastable state in which the system can be dynamically trapped.

\section{Dynamics of self-organization}
\label{sec:dynamics}

We now examine the dynamics of the system when the values of $\alpha_1$ and $\alpha_2$ are varied as a function of time. Experimentally, this corresponds to vary the pump laser intensities or their detuning with respect to the cavity mode frequencies. At time $t=0$ we assume that the system is prepared in the stationary state of a paramagnetic phase, described by the distribution in Eq.\ \eqref{f:st} by setting $\alpha_n=\alpha_{ni}\ll 1$ in Eq.\ \eqref{eq:hamiltonian} ($n=1,2$). The values $\alpha_n$ appearing in the equations of motion \eqref{com} are then varied in time, by performing either (i) a sudden quench, i.e.\ suddenly switching the two values of $\alpha_{1f}$ and $\alpha_{2f}$, or (ii) a slow quench, consisting in varying $\alpha_n(t)$ monotonously and continuously in time towards the final values $\alpha_{1f}$ and $\alpha_{2f}$. We choose the final values $\alpha_{nf}$ in the ferromagnetic phase. The quench protocols we consider are illustrated by the green lines in Fig.\ \ref{fig:phase_diagram}(a): for sudden quenches, the initial and final values are two points connected by the green line. A slow quench sweeps across the intermediate points along the line. We are interested in determining the dynamics leading to the steady state.

%The central question we address here is, how well the final meta-stable state after the ramp coincides with the equilibrium solutions of Eq. \eqref{f:st} and how long it takes afterwards until the stationary equilibrium is reached, which is expected at the asymptotic time limit of the dynamics following the quench according to the predictions of the semi-classical equations \eqref{com}, see Ref. \cite{Keller:2017}. 

In what follows we perform numerical simulations of Eq.\ \eqref{com} using the parameters of a gas of $^{85}$Rb atoms. In particular, we take $k=2\pi/\lambda$ with $\lambda=780 \mathrm{nm}$ the wavelength of the $D_2$ line. The corresponding recoil frequency is $\omega_r=2\pi\times3.86 \mathrm{kHz}$. The cavity linewidth is taken to be $\kappa=2\pi\times1.5 \mathrm{MHz}$, so that  $\kappa\approx388.6\omega_r$. A possible realization of the two-mode setup here considered has been discussed in Refs. \cite{Keller:2017,Leonard:2017}.

\subsection{Sudden quench into the ferromagnetic phase}
\label{subsec:quench_ferro}

We first consider sudden quenches from $\alpha_{1i}, \alpha_{2i}$ in the paramagnetic phase to $\alpha_{1f}, \alpha_{2f}$ in the ferromagnetic phase, keeping $\alpha_{1i}/\alpha_{2i} = \alpha_{1f}/\alpha_{2f}=5$. The initial values are vanishingly small and the atoms are at the corresponding stationary distribution, which is a thermal distribution at the temperature determined by the corresponding detuning, Eq. \eqref{T}, with homogeneous density.  The detuning before and after the quench is taken to be equal, thus it is expected that the atoms reach a thermal distribution with the same temperature as the initial state. 

\begin{figure}
\centering
\includegraphics[width=0.49\textwidth]{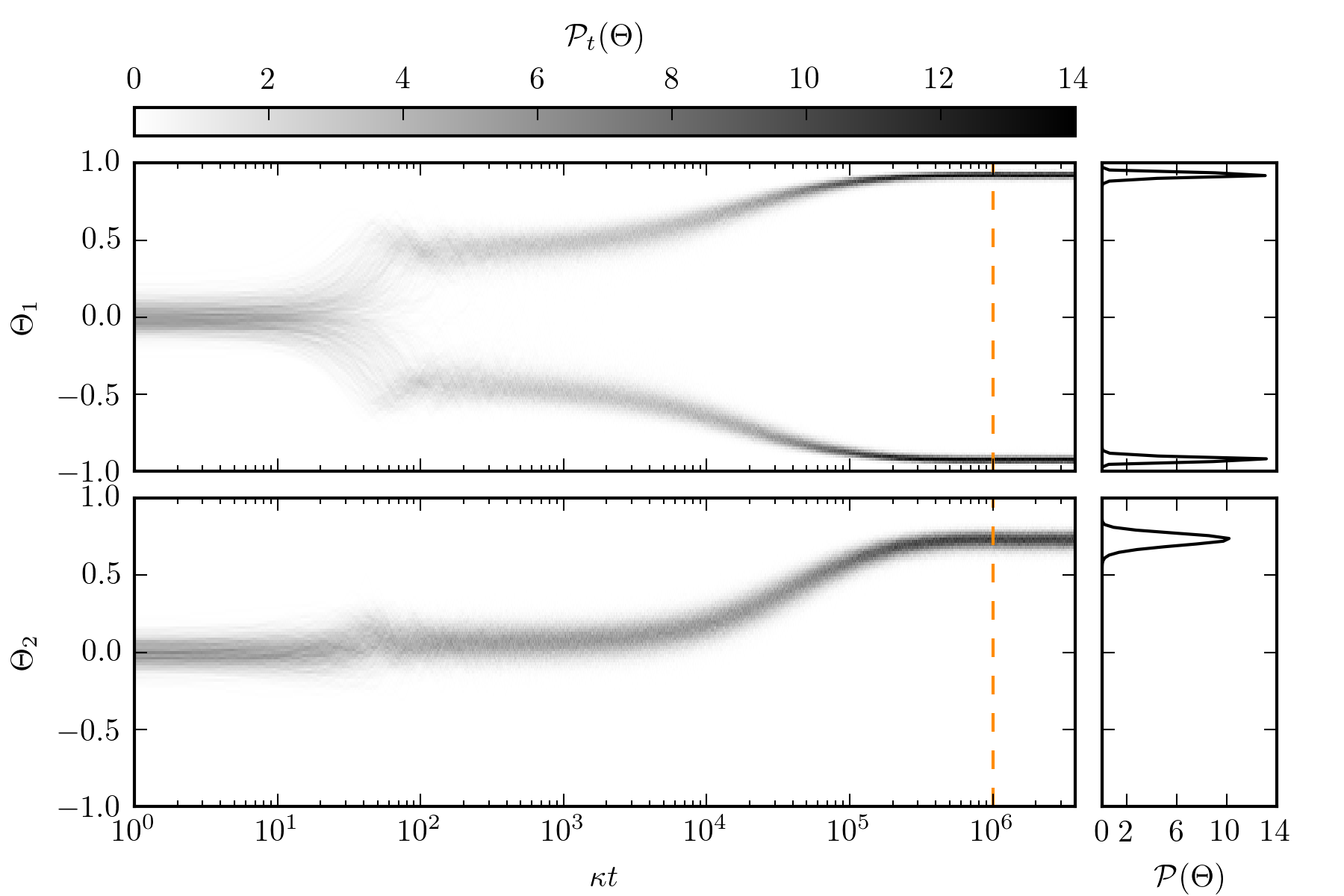}
\caption{Dynamics following a sudden quench from $\alpha_{1i},\alpha_{2i}\ll 1$ to $\alpha_{1f}=2.5$, $\alpha_{2f}=0.5$ keeping constant the detuning $\Delta_c=-\kappa$. The left panel displays the contour plot of distribution $\mathcal P_t(\Theta)$ (Eq.\ \eqref{P:t:Theta}) for $\Theta=\Theta_1$ and $\Theta=\Theta_2$ as a function of time (in units of $1/\kappa$). The distribution has been extracted from the numerical simulations using Eq. \eqref{com} for $N=100$ atoms and $1000$ trajectories. The grid in $\Theta$ for the left panel has minimum step $\Delta_\Theta=2/111$, the grey scale gives the relative weight. The right panels display the distributions $\mathcal P(\Theta_1)$ and $\mathcal P(\Theta_2)$ as a function of $\Theta_1$ and $\Theta_2$, respectively, see Eq. \eqref{P:Theta}. Here, the time average has been performed for $N_t=113$ instants of time chosen between $t_1=10^6/\kappa$ and $t_{\mathrm{f}}=3.77\times 10^6/\kappa$. The vertical dashed line in the left panels marks the instant of time $t_1=10^6/\kappa$.  }
\label{fig:3}
\end{figure}

Figure \ref{fig:3} displays the distribution $\mathcal P_t(\Theta)$ for the order parameters $\Theta_1$ and $\Theta_2$ as a function of time for $(\alpha_{1f}, \alpha_{2f}) = (2.5, 0.5)$. It is defined as a time sequence of normalized histograms
\begin{equation}
\mathcal P_t(\Theta)=\frac{\text{\# trajectories with $\Theta(t) \in [\Theta-\Delta_\Theta/2,\Theta+\Delta_\Theta/2]$}}{\text{\# trajectories}\times \Delta_{\Theta}}\, ,
\label{P:t:Theta}
\end{equation}
where $\Theta$ is calculated on each trajectory of the simulations with the stochastic different equations and its value is determined according to the precision $\Delta_\Theta$ of the grid in $\Theta$. We observe that at a given time scale of the order of $10^2/\kappa$, $\mathcal P_t(\Theta_1)$ splits into two branches corresponding to two possible orders in the long-wavelength lattice. This symmetry breaking is well known from the single mode case \cite{Asboth:2005}. The order parameter of the short-wavelength mode $\Theta_2$, which is weakly pumped, substantially grows to a positive value long after the symmetry breaking. The fact that $\mathcal P_t(\Theta_2)$ vanishes for negative $\Theta_2$ values comes from the ordering of the atoms close to the anti-nodes of the dominant long-wavelength mode field $\cos(kx)$ (see Fig.\ \ref{Fig:1}).

%Ordering in the first, long-wavelength lattice acts as a bias-field for the second lattice, where the values of $\Theta_2$ become positive long after the symmetry breaking transition for $\Theta_1$ has taken place. This asymmetry in $\Theta_1$ comes from the ordering of the atoms close to the anti-nodes of the dominant long-wavelength mode field $\cos(kx)$, where in our set-up the short-wavelength mode field $\cos(2kx)$ is positive.

The distributions $\mathcal P(\Theta_n)$ at the asymptotics are reported in the right panels of Fig.\ \ref{fig:3}. They are obtained by averaging $\mathcal{P}_t(\Theta_n)$ over times $t\ge 10^6/\kappa$, where a stable configuration has been reached. Formally
\begin{eqnarray}
\label{P:Theta}
\mathcal P(\Theta)=\sum_{i=1}^{N_t} \mathcal P_{t_i}(\Theta)/N_t\,,
\end{eqnarray}
where $N_t$ is the number of instants of times at which the distribution is sampled in the interval $[t_1,t_{\mathrm{f}}]$, with $t_1=10^6/\kappa$ and $t_{\mathrm{f}}=t_{N_t}>t_1$. 
Comparing the widths of the distributions in the right panel of Fig. \ref{fig:3} one observes that after sufficiently long times the long-wavelength order parameter fluctuates less than the short-wavelength order parameter. 

\begin{figure}[h!]
\flushleft(a)\\
\center\includegraphics[width=0.4\textwidth]{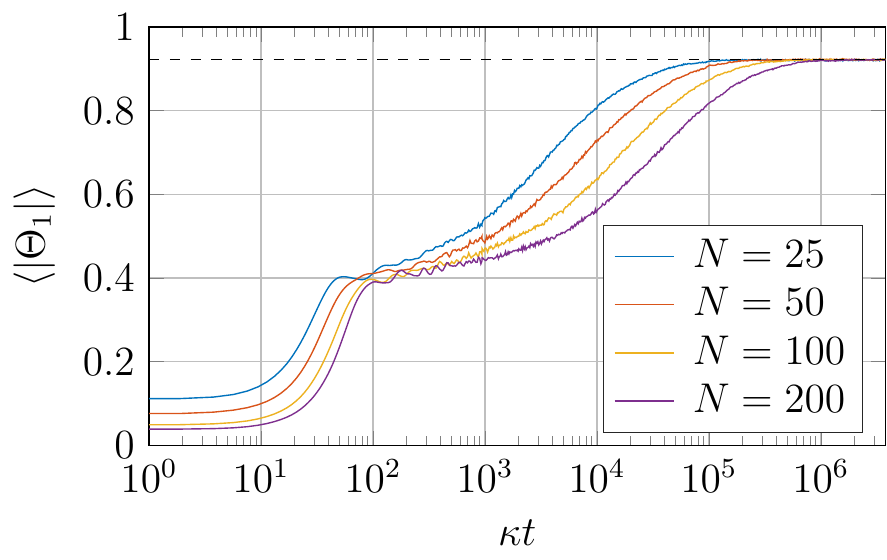}
\flushleft(b)\\
\center\includegraphics[width=0.4\textwidth]{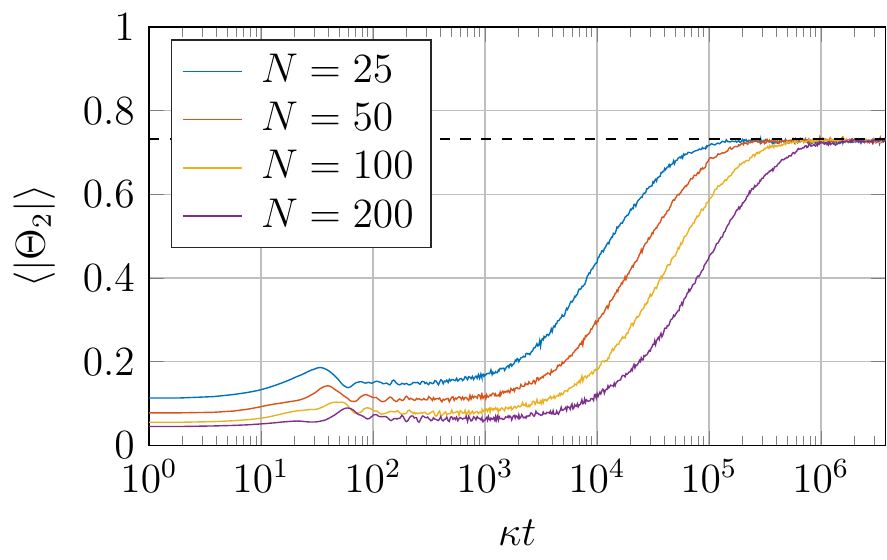}
\flushleft(c)\\
\center\includegraphics[width=0.37\textwidth]{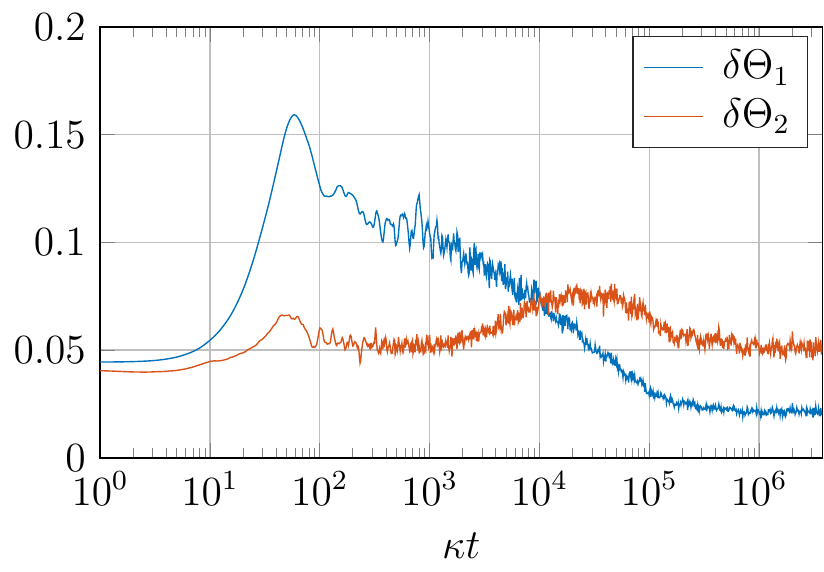}
\caption{Dynamics of (a) $\langle|\Theta_1|\rangle$, (b) $\langle|\Theta_2|\rangle$, and (c) their fluctuations $\delta\Theta_n=\sqrt{\langle\Theta_n^2\rangle-\langle|\Theta_n|\rangle^2}$ as a function of time (in units of $1/\kappa$). The parameters and quench protocol are the same as in Fig. \ref{fig:3}, the curves are however evaluated for different numbers of atoms and of trajectories. In (a) and (b) the data correspond to $N=25,50,100,200$ particles (see legenda for color code) and respectively $1000,500,250,125$ trajectories. The horizontal dashed lines indicate the value predicted by the global minimum of the free energy in Eq.\eqref{freeenergy}. The finite values of the order parameters at $t=0$ are due to finite size effects, $\langle |\Theta_n(0)|\rangle=1/\sqrt{\pi N}$.  The curves in (c) are calculated for $N=100$ and $250$ trajectories.}
\label{fig:ferro_quench:Theta}
\end{figure}

Figures \ref{fig:ferro_quench:Theta}(a)-(b) display the dynamics of the mean absolute value of the order parameters for different values of $N$. Figure \ref{fig:ferro_quench:Theta} (c) shows the time evolution of the fluctuations of the order parameters $\delta\Theta_n=\sqrt{\langle\Theta_n^2\rangle-\langle|\Theta_n|\rangle^2}$ for $N=100$ particles. The order parameters asymptotically tend to the values predicted by the free energy, indicated by the horizontal dashed line, for a time scale of the order of $10^6/\kappa$. Meanwhile the fluctuation $\delta\Theta_1$ relaxes to a much smaller value than the asymptotic value of $\delta\Theta_2$ reproducing the widths of the distributions in the right panel in Fig. \ref{fig:3}. The time evolution of $\langle |\Theta_1|\rangle$, in particular, is reminiscent of the one observed for quenches into the ferromagnetic phase in a single-mode resonator \cite{Schuetz:2016}. It can be separated into three stages which we denote by (in order of their temporal appearance) (i) violent relaxation, corresponding to an exponential increase of the absolute value of the order parameter $\langle|\Theta_1|\rangle$; (ii) transient dynamics, corresponding to power-law scaling with time, and (iii) relaxation phase, where the mean values tend exponentially towards the asymptotic value. The violent relaxation can be described by a mean-field model \cite{Jaeger:2016}, in the transient stage the coherent dynamics is prevailing, while the relaxation stage is dominated by dissipation \cite{Schuetz:2016}. The transient and relaxation stages are characterized by time scales which increase with $N$ but with different functional dependence \cite{Jaeger:2016}. The time scale $10^6/\kappa$ can here be identified as the one at which the asymptotic state is reached for $N\lesssim 200$, while for larger numbers of particles longer time scales shall be considered. 

%
%Therefore during this time interval the phase is nematic. However, this is not the nematic phase that is predicted by the phase diagram, see Fig. \ref{fig:phase_diagram} (a), and is not a stationary state of the system. Yet, it is metastable, and for $N\leq 200$ particles its lifetime is of the order of $t\sim10^4/\kappa$. From Fig. \ref{fig:ferro_quench:Theta} (b), we expect that the lifetime can further grow as $N$ increases.

\begin{figure}[h!]
\flushleft(a)\\
\center\includegraphics[width=0.4\textwidth]{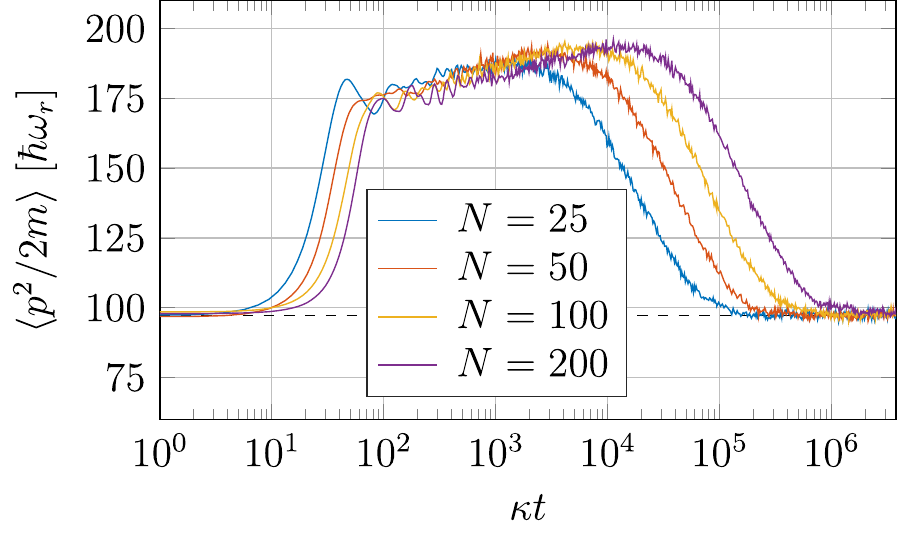}
\flushleft(b)\\
\center\includegraphics[width=0.4\textwidth]{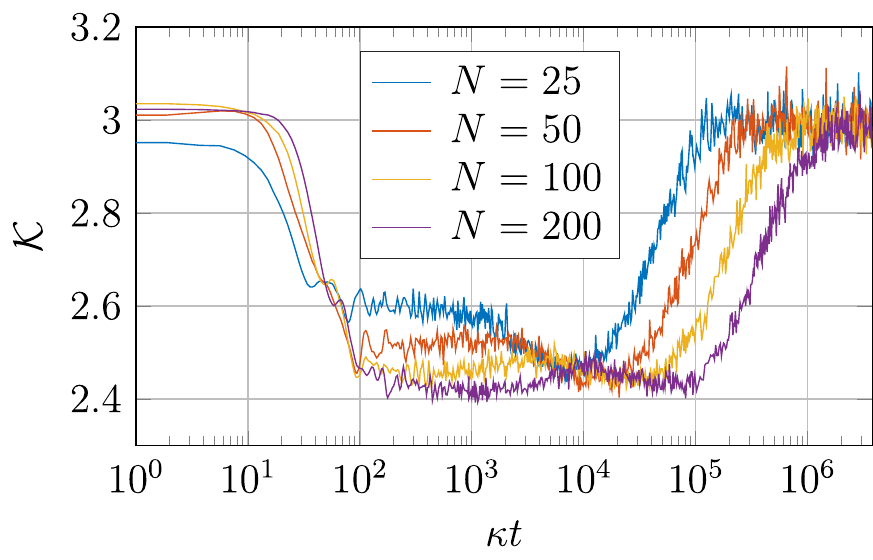}
\caption{Dynamics of (a) the single-particle kinetic energy $\langle p^2/2m\rangle$ (in units of $\hbar\omega_r$) and (b) the Kurtosis $\mathcal K=\langle p^4\rangle/\langle p^2\rangle^2$, for $N=25,50,100,200$ particles (see legenda) and correspondingly $1000,500,250,125$ trajectories. The other parameters and quench protocol are the same as in Fig. \ref{fig:ferro_quench:Theta}. The horizontal dashed line in (a) indicates the asymptotic value predicted by Eq. \eqref{T}.}
\label{fig:ferro_quench:p}
\end{figure}

Interestingly, in the transient phase there is ordering only in the long-wavelength mode of the cavity, while ferromagnetic order  is finally established by dissipation on a longer timescale. The metastable phase of the transient dynamics can be therefore denoted by ''nematic'', its lifetime increases with $N$ and for $N\sim 200$ it is of the order of $t\sim10^4/\kappa$. However, this metastable ''nematic'' state cannot be understood in terms of the landscape of the free energy, but rather seems to exhibit the features of the quasi-stationary state due to the long-range coherent dynamics analogous to the ones reported in Ref. \cite{Teles:2012}. This conjecture is also supported by the behaviour of the  single-particle kinetic energy and of the kurtosis $\mathcal K=\langle p^4\rangle/\langle p^2\rangle^2$, which are shown in Fig. \ref{fig:ferro_quench:p}. The latter quantifies the deviation of the momentum distribution from a Gaussian, for which it takes the value $\mathcal K_{\rm Gauss}=3$. For these quantities we observe that in the metastable nematic phase the kinetic energy grows, while the distribution is non-thermal. Ordering in the second, short-wavelength lattice is accompanied by cooling into a thermal distribution.

\begin{figure}[h!]
\centering
\includegraphics[width=0.4\textwidth]{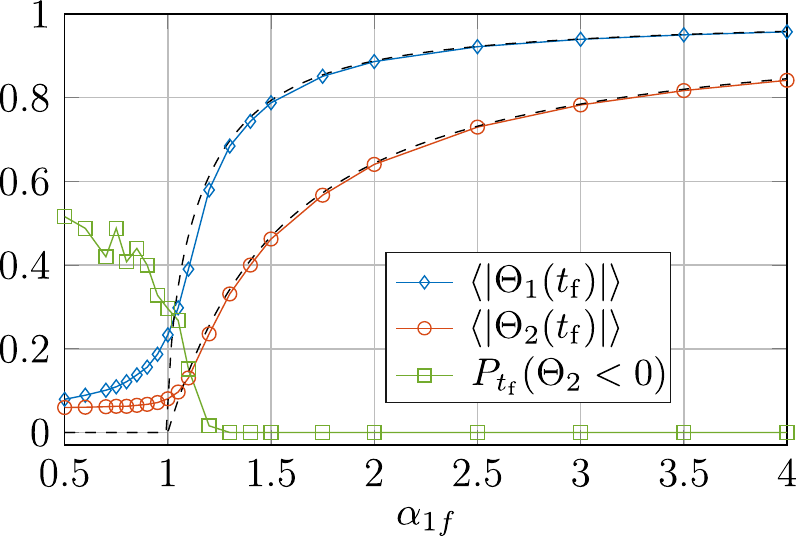}
\caption{Asymptotic values of $\langle |\Theta_1(t_{\mathrm{f}})|\rangle$, $\langle |\Theta_2(t_{\mathrm{f}})|\rangle$, and $P_{t_\mathrm{f}}(\Theta_2<0)$, Eq. \eqref{eq:def_p2}, as a function of $\alpha_{1f}$. The quenches start all from the same initial values in the paramagnetic phase ($\alpha_{1i},\alpha_{2i}\ll1$ and $\Delta_c=-\kappa$) and end up in different values $\alpha_{1f},\alpha_{2f}$ with $\alpha_{1f}=5\alpha_{2f}$ (lying along  the left green line in Fig. \ref{fig:phase_diagram}(a)) and $\Delta_c=-\kappa$. The circles correspond to the results of the numerical simulations at $t_{\mathrm{f}}=3.77\times 10^6/\kappa$ with $N=100$ particles and $250$ trajectories. The dashed lines indicate the predictions of the global minima of Eq. \eqref{freeenergy}.}
\label{fig:ferro_quench:steady}
\end{figure}

We now compare the numerical results with the analytic theory for different quenches starting from the same initial values of $\alpha_{1i},\alpha_{2i}\ll1$ but with different endpoints $\alpha_{1f},\alpha_{2f}$. We take different endpoints ranging from the paramagnetic to the ferromagnetic phase, under the constrain $\alpha_{1f}/\alpha_{2f}=5$. The circles in Fig. \ref{fig:ferro_quench:steady} correspond to the numerical results for 100 particles at time $t_{\mathrm{f}}>10^6/\kappa$, where we expect that the system has reached the steady state. These are in good agreement with the analytical results (dashed lines) based on evaluating the corresponding observables at the global minimum of the free energy. The interval where $\langle |\Theta_n|\rangle$ grows monotonically from $\sim 1/\sqrt{N}$ to the value of the ferromagnetic phase is expected to shrink as $N$ is increased, in agreement with a second order phase transition at the thermodynamic limit. 
Further information on the onset of this ferromagnetic order can be gained by the probability $P_t(\Theta_2<0)$ that $\Theta_2$ is negative at $t$:
\begin{equation}
P_t(\Theta_2<0)=\int_{-1}^0{\rm d}\Theta_2\mathcal P_t(\Theta_2)\, .
\label{eq:def_p2}
\end{equation}
We note that in the paramagnetic phase (homogeneous spatial distribution) we expect $P_{t}(\Theta_2<0)\simeq 0.5$. In contrast, due to the given mode structure we expect that $P_t(\Theta_2<0) \simeq 0$ for long-wavelength ordering in the ferromagnetic phase. Indeed, as $\alpha_{1f}$ increases across the critical value, $P_t(\Theta_2<0)$, quickly drops down to zero.

\subsection{Sudden quenches into the bistable phase}
\label{subsec:quench_mixed}

We now turn to the dynamics following sudden quenches from the paramagnetic to the ferromagnetic phase but following the right path of Fig. \ref{fig:phase_diagram}(a), which consists in equal effective pumping $\alpha_{1i}/\alpha_{2i} = \alpha_{1f}/\alpha_{2f}=1$. In this parameter region (bistable phase) the free energy exhibits a local minimum, which is nematic. As in the previous case, the initial values $\alpha_{1i},\alpha_{2i} $ are vanishingly small and the atoms are at the corresponding stationary distribution, whose temperature is determined by the detuning $\Delta_c$ and whose spatial density is homogeneous. The quench is performed by switching the laser power keeping the detuning constant, thus the atoms should reach a thermal distribution with the same temperature as the initial state.

\begin{figure}[h!]
\centering
\includegraphics[width=0.49\textwidth]{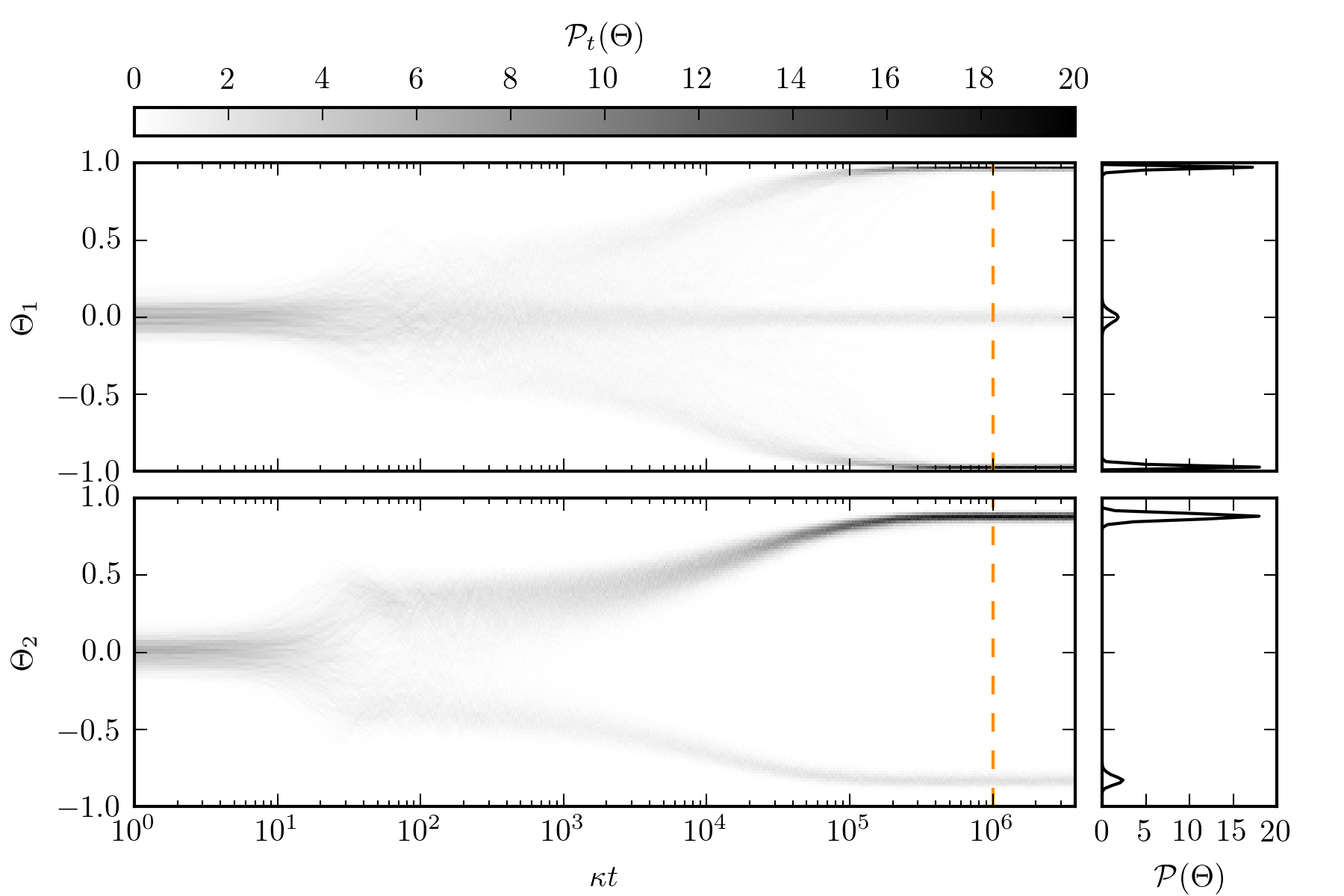}
\caption{
Dynamics following a sudden quench from $\alpha_{1i},\alpha_{2i}\ll 1$ to $\alpha_{1f}=\alpha_{2f}=2$ keeping constant the detuning $\Delta_c=-\kappa$ (red circle (c) in Fig. \ref{fig:phase_diagram}(a)). The left panel displays the contour plot of the distribution $\mathcal P_t(\Theta)$ (Eq.\ \eqref{P:t:Theta}) for $\Theta=\Theta_1$ and $\Theta=\Theta_2$ as a function of time (in units of $1/\kappa$). The distribution has been extracted from the numerical simulations using Eq. \eqref{com} for $N=100$ atoms and $1000$ trajectories. The right panels display the distributions $\mathcal P(\Theta_1)$ and $\mathcal P(\Theta_2)$ as a function of $\Theta_1$ and $\Theta_2$, respectively, see Eq. \eqref{P:Theta}. See Fig. \ref{fig:3} for further details.
}
\label{fig:mixed_quench}
\end{figure}

\begin{figure}[h!]
\flushleft(a)\\
\center\includegraphics[width=0.4\textwidth]{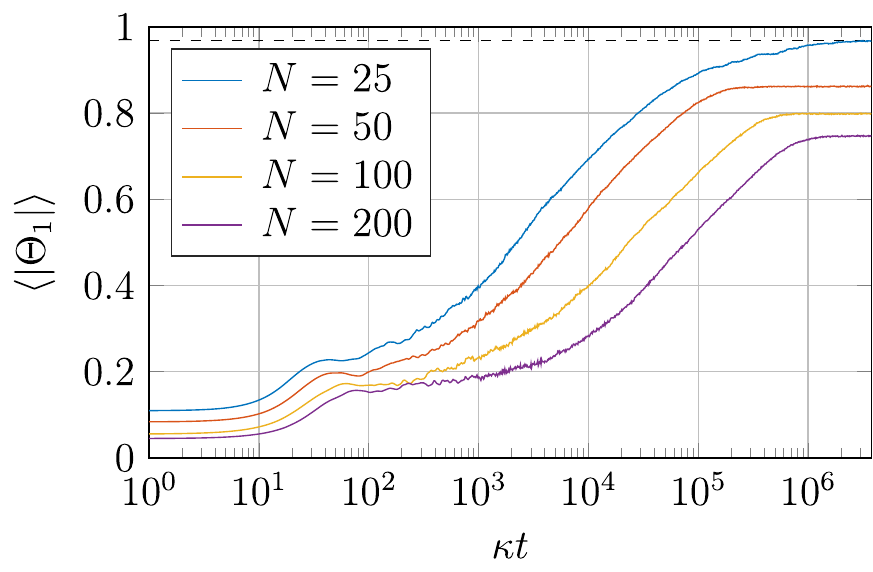}
\flushleft(b)\\
\center\includegraphics[width=0.4\textwidth]{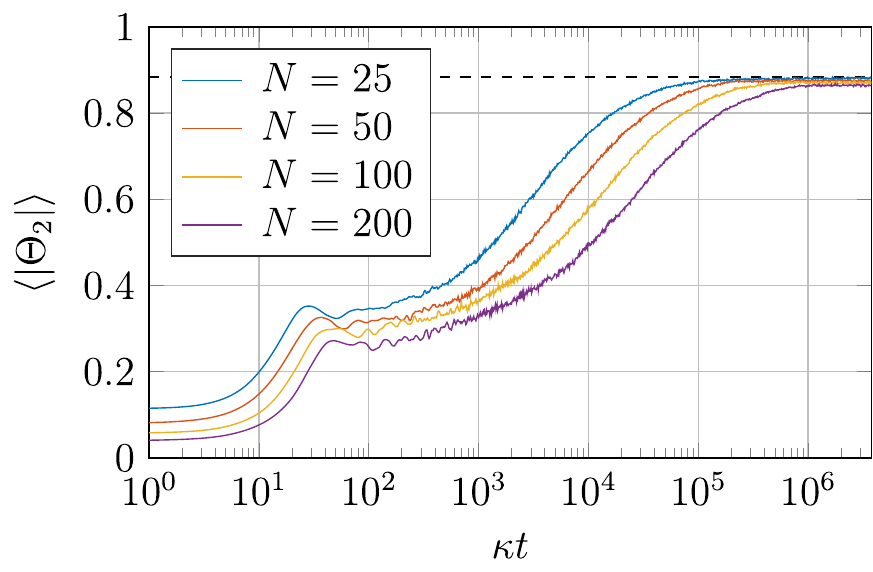}
\flushleft(c)\\
\center \includegraphics[width=0.37\textwidth]{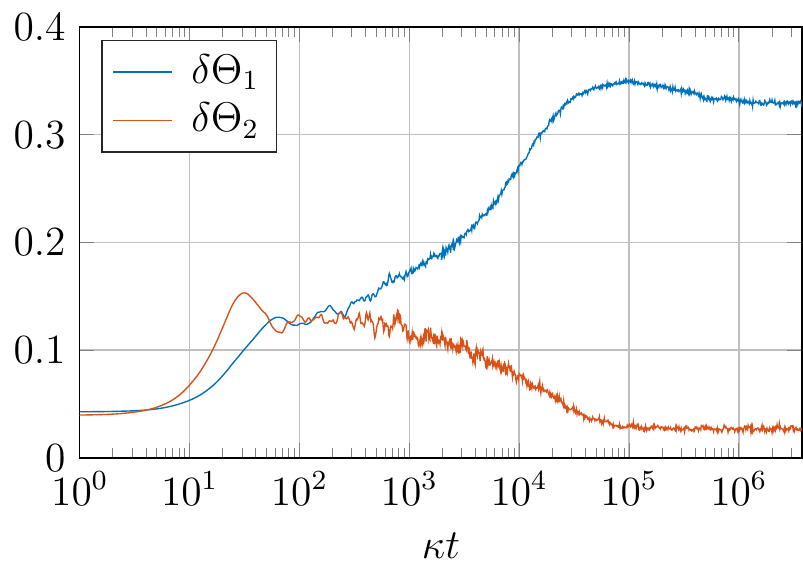}
\caption{
Dynamics of (a) $\langle|\Theta_1|\rangle$, (b) $\langle|\Theta_2|\rangle$, and (c) their fluctuations $\delta\Theta_n=\sqrt{\langle\Theta_n^2\rangle-\langle|\Theta_n|\rangle^2}$ as a function of time (in units of $1/\kappa$). The parameters and quench protocol are the same as in Fig. \ref{fig:mixed_quench}, the curves are evaluated for different numbers of atoms and of trajectories. In (a) and (b) the data correspond to $N=25,50,100,200$ particles (see legenda for color code) and respectively $1000,500,250,125$ trajectories. The horizontal dashed lines indicate the value predicted by the global minimum of the free energy in Eq. \eqref{freeenergy}. The finite values of the order parameters at $t=0$ are due to finite size effects, $\langle |\Theta_n(0)|\rangle=1/\sqrt{\pi N}$.  The curves in (c) are calculated for $N=100$ and $250$ trajectories.
}
\label{fig:mixed_quench:Theta}
\end{figure}

Figure \ref{fig:mixed_quench} displays the time evolution of the trajectories' $\Theta$-distribution for $\alpha_{1f}=\alpha_{2f}=2$ and $\Delta_c=-\kappa$. As opposed to the previous section, here a finite fraction of trajectories gets trapped in the nematic phase with vanishing value of $\Theta_1$ and finite probability that $\Theta_2$ takes negative values. This is visible in the small extra peaks in $\mathcal{P}(\Theta_1)$ and $\mathcal{P}(\Theta_2)$ (right panels). The trapping occurs at the time scale of the violent relaxation, and it seems stable over times of the order of $10^6/\kappa$. We conjecture that it persists also at asymptotic times. In Fig.\ \ref{fig:mixed_quench:Theta} the time evolution of the mean absolute value of the order parameters is shown for different numbers of particles. While $\langle |\Theta_2|\rangle$ reaches the same stationary value (in reality its value decreases slightly with $N$), instead the asymptotic value of $\langle |\Theta_1|\rangle$ decreases as $N$ grows. This suggests that the probability that the dynamics gets trapped in the local minimum increases with the number of particles. The asymptotic value of $\delta\Theta_1=\sqrt{\langle\Theta_1^2\rangle-\langle|\Theta_1|\rangle^2}$ in subplot (c) reflects the contribution of these trajectories.

\begin{figure}[h!]
\flushleft(a)\\
\center \includegraphics[width=0.4\textwidth]{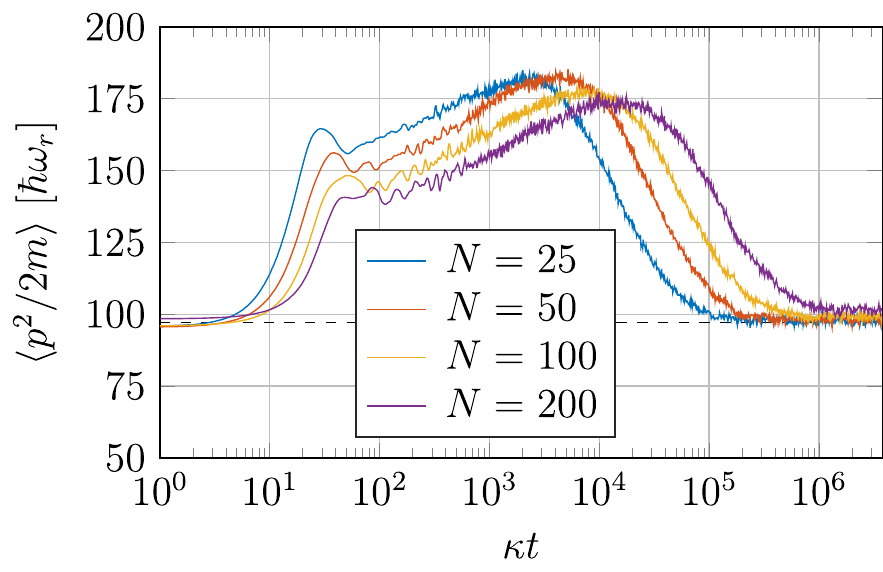}
\flushleft(b)\\
\center\includegraphics[width=0.4\textwidth]{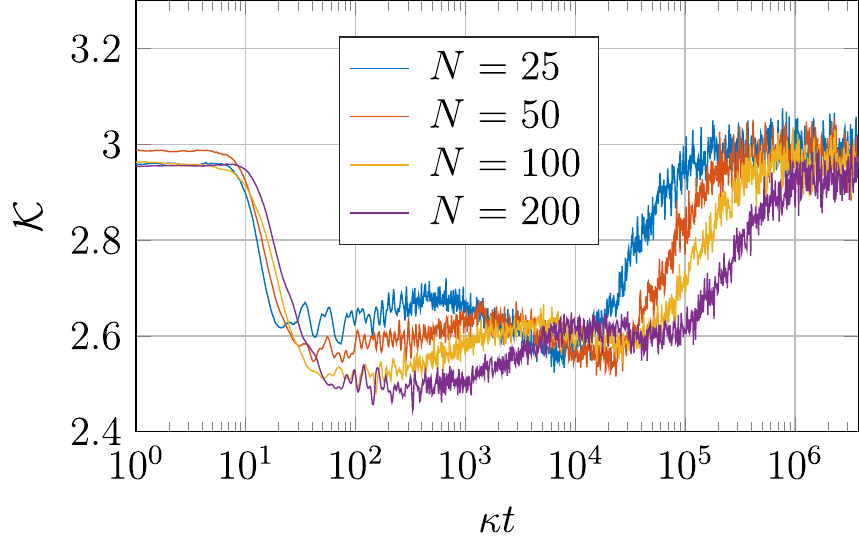}
\caption{
Dynamics of (a) the single-particle kinetic energy $\langle p^2/2m\rangle$ (in units of $\hbar\omega_r$) and (b) the Kurtosis $\mathcal K=\langle p^4\rangle/\langle p^2\rangle^2$, for $N=25,50,100,200$ particles (see legenda) and correspondingly $1000,500,250,125$ trajectories. The other parameters and initial conditions are the same as in Fig. \ref{fig:mixed_quench:Theta}. The horizontal dashed line in (a) indicates the asymptotic value predicted by Eq. \eqref{T}.
}
\label{fig:mixed_quench:p}
\end{figure}

Mean single-particle kinetic energy and kurtosis are shown in Fig. \ref{fig:mixed_quench:p}. From their behaviour we infer that the metastable nematic state does not significantly deviate from a thermal distribution with the expected asymptotic temperature (Eq.\ \eqref{T}). %This metastable state is consistent with the properties of the local minimum of the free energy, Eq. \eqref{freeenergy}, in the bistable region meaning that a part of the atoms form a nematic order, which locally minimize the free energy and is at the asymptotic temperature $k_BT$, Eq. \eqref{T}. 

%The probability that the dynamics gets trapped in metastable states increases with the number of particles, as visible in Fig. \ref{fig:mixed_quench:Theta}. As $N$ grows, in fact, the asymptotic value of $\langle |\Theta_1|\rangle$ decreases, see Fig. \ref{fig:mixed_quench:Theta}(a). The three stage dynamics can be observed also in this case. However, while $\langle |\Theta_2|\rangle$ approaches the asymptotic value predicted by the free energy, $\langle |\Theta_1|\rangle$ tends to smaller values due to the contribution of an increasing number trajectories which asymptotically display $\langle |\Theta_1|\rangle\sim 0$. This can also be seen in the dynamics of the fluctuations in subplot (c): The asymptotic value of $\delta\Theta_1=\sqrt{\langle\Theta_1^2\rangle-\langle|\Theta_1|\rangle^2}$ reflects the contribution of these trajectories.

\begin{figure}[h!]
\centering
\includegraphics[width=0.4\textwidth]{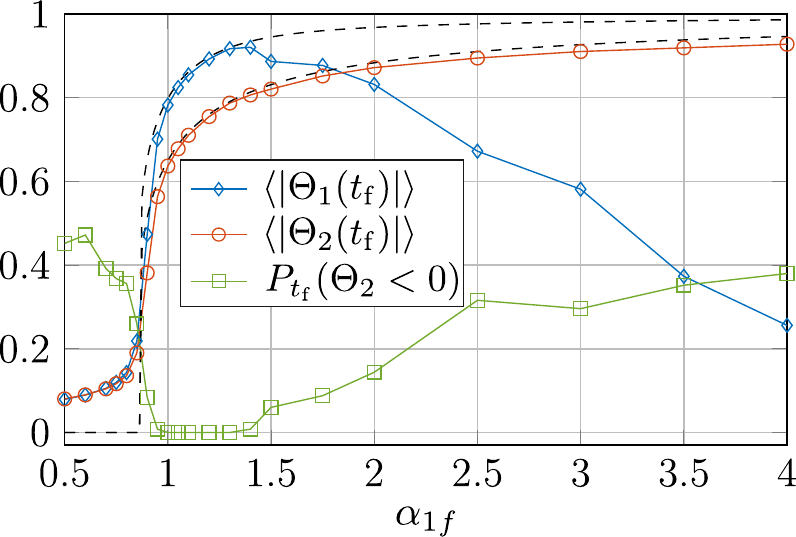}
\caption{
Asymptotic values of $\langle |\Theta_1(t_{\mathrm{f}})|\rangle$, $\langle |\Theta_2(t_{\mathrm{f}})|\rangle$, and $P_{t_\mathrm{f}}(\Theta_2<0)$, Eq. \eqref{eq:def_p2}, as a function of $\alpha_{1f}$. The quenches start all from the same initial values in the paramagnetic phase ($\alpha_{1i},\alpha_{2i}\ll1$ and $\Delta_c=-\kappa$) and end up in different values $\alpha_{1f},\alpha_{2f}$ with $\alpha_{1f}=\alpha_{2f}$ (lying along  the right green line in Fig. \ref{fig:phase_diagram}(a)) and $\Delta_c=-\kappa$. The circles correspond to the results of the numerical simulations at $t_{\mathrm{f}}=3.77\times 10^6/\kappa$ with $N=100$ particles and $250$ trajectories. The dashed lines indicate the predictions of the global minima of Eq. \eqref{freeenergy}.}
\label{fig:mixed_quench:Steady}
\end{figure}

Peculiar features of these dynamics become visible when inspecting the probability $P_{t}(\Theta_2<0)$ at the asymptotics and as a function of $\alpha_{1f}$ in Fig.\ \ref{fig:mixed_quench:Steady}. As in Fig.\ \ref{fig:ferro_quench:steady}, it vanishes identically upon leaving the paramagnetic phase, but increases again as $\alpha_{1f},\alpha_{2f}$ are chosen deeper in the bistable phase of Fig.\ \ref{fig:phase_diagram}(a). Correspondingly, $\langle |\Theta_1|\rangle$ starts to decrease as $\alpha_{1f}$ increases, which suggests that from this point on the depth of the local minimum grows. The value of the order parameter  $\langle |\Theta_2|\rangle$ at which $P_{t}(\Theta_2<0)$ starts to grow again identifies a threshold, above which the local minimum is sufficiently deep to stably trap particles.

\subsection{Slow ramp into the bistable phase}
\label{subsec:slowramp}

We now consider linear ramps of $\alpha_n(t)$ across the transition region separating the paramagnetic from the bistable region. The ramps protocols have duration $\tau$ and sweep between the values $[\varepsilon,\alpha_{nf}]$, with $\varepsilon\ll 1$. In particular, $\alpha_n(t)=\varepsilon+\alpha_{nf}\frac{t}{\tau}$ if $t\in[0,\tau]$, while for $t>\tau$ then $\alpha_n(t)$ is constant and equal to $\alpha_{nf}$. Note that a sudden quench is the limit $\tau\to 0$ of a linear quench. We choose to vary the values of $\alpha_n(t)$ along the rightmost green line in Fig.\ \ref{fig:phase_diagram}(a), so that $\alpha_1(t)=\alpha_2(t)$ at all instants of time, with $\alpha_{nf}$ in the bistable phase. We further keep $\Delta_c$ constant, and solely vary the pump intensity. This means that the asymptotic temperatures at each value of $\alpha_n$ are equal.

\begin{figure}[h!]
\flushleft(a)\\
\center \includegraphics[width=0.4\textwidth]{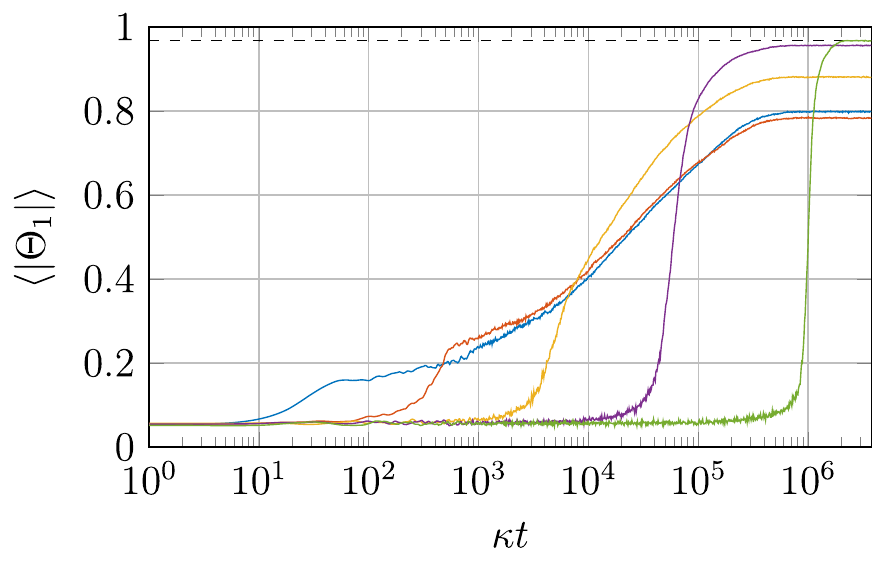}
\flushleft(b)\\
\center\includegraphics[width=0.4\textwidth]{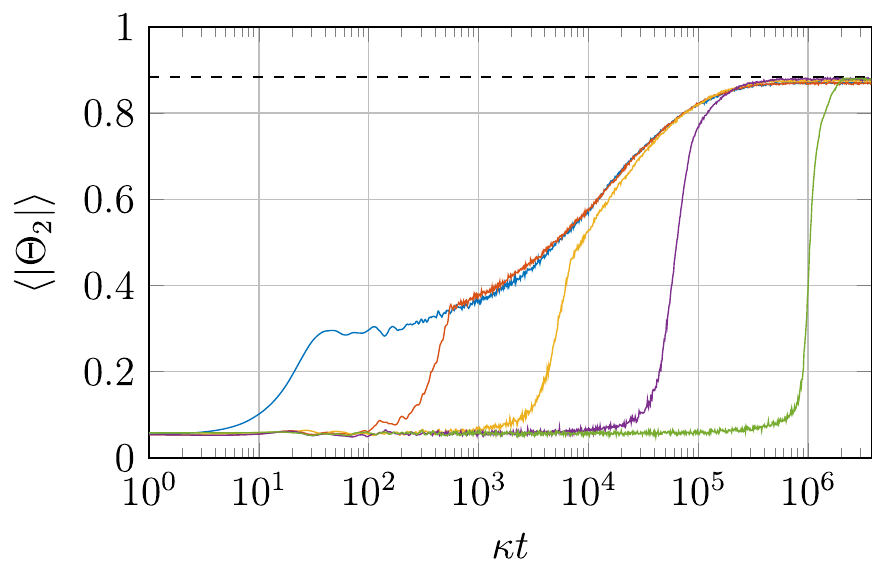}
\caption{Mean value of the order parameters, (a) $\langle|\Theta_1|\rangle$ and (b) $\langle|\Theta_2|\rangle$ as a function of time (in units of $1/\kappa$) for $N=100$ and $\Delta_c=-\kappa$, evaluated numerically with 250 trajectories. The curves are the time evolution during and after linear ramps of duration $\tau=0$ (blue); $\tau=5.5\times 10^2/\kappa$ (red),  $\tau=6.8\times 10^3/\kappa$ (yellow), $\tau=8.5\times 10^4/\kappa$ (purple), $\tau=2\times 10^6/\kappa$ (green). The ramps are from the paramagnetic to the bistable phase, specifically, from $\alpha_{1i}=\alpha_{2i}=\varepsilon\ll 1$ to  $\alpha_{1f}=\alpha_{2f}=2$. As before, at $t=0$ the initial state of the atoms is the steady state, Eq. \eqref{f:st}, for $\alpha_n=\alpha_{ni}$ and $\Delta_c=-\kappa$. The dashed horizontal lines show the steady state values of the global minima of the free energy, Eq. \eqref{freeenergy}. 
}
\label{fig:ramp:Theta}
\end{figure}

Figure \ref{fig:ramp:Theta} displays the dynamics of the mean absolute value of the order parameters for $\alpha_{1f}=\alpha_{2f}=2$ for linear ramps with different durations $\tau$. The dynamics following the sudden quench (cif.\ Fig.\ \ref{fig:mixed_quench:Theta} (a) and (b)) is shown for comparison (blue curve). We observe that the dynamics of the order parameters exhibits an exponential increase which occurs almost simultaneously for both $\langle|\Theta_1|\rangle$ and $\langle|\Theta_2|\rangle$. This behaviour seems to be initiated at the instant of time when the parameters $\alpha_n(t)$ cross the critical point of the phase diagram. Moreover, for sufficiently slow ramps $\langle |\Theta_1|\rangle$ approaches the asymptotic value of the free energy's global minimum, signaling stationary long-wavelength order.

\begin{figure}[h!]
	\centering
	\includegraphics[width=0.4\textwidth]{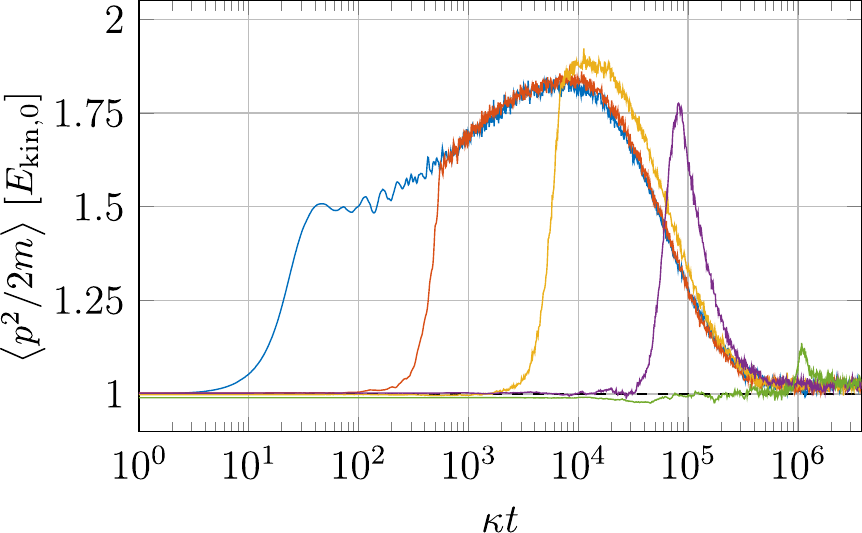}
	\caption{Mean value of the single-particle kinetic energy, $\langle p^2/(2m)\rangle$ (in units of $E_{\mathrm{kin},0}=\hbar \kappa/4$) as a function of time (in units of $1/\kappa$) for the same parameters and color codes as in Fig. \ref{fig:ramp:Theta}.
	}
	\label{fig:ramp:p}
\end{figure}

\begin{figure}[h!]
	\centering
	\includegraphics[width=0.4\textwidth]{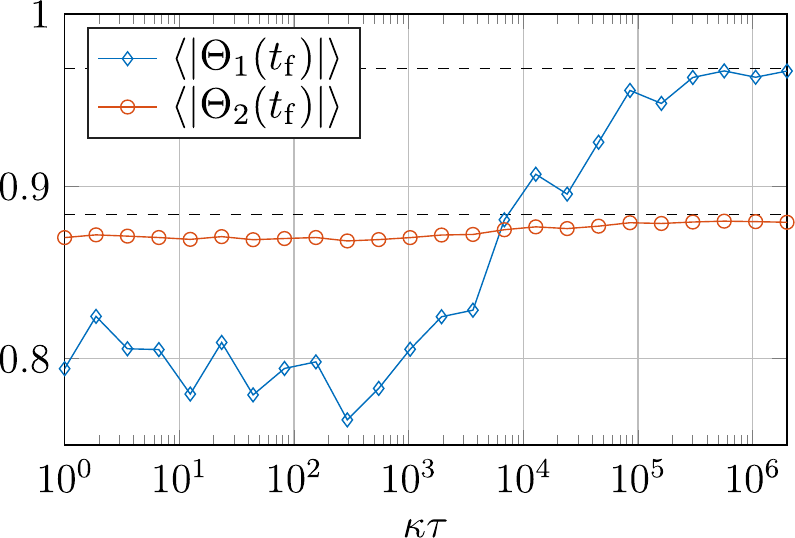}
	\caption{Values of  $\langle |\Theta_1(t_{\mathrm{f}})|\rangle$ (blue) and $\langle |\Theta_2(t_{\mathrm{f}})|\rangle$ (red) at $t_{\mathrm{f}}=3.77\times 10^6/\kappa$ and as a function of the ramp duration $\tau$ (in units of $1/\kappa$), for the same parameters as in Fig. \ref{fig:ramp:Theta}. The dashed horizontal lines show the steady state value predicted by the free energy, Eq. \eqref{freeenergy}. 
	}
	\label{fig:ramp:St}
\end{figure}

We further note that for $\tau\lesssim 10^3/\kappa$ the orders parameters undergo a three-stage dynamics, as for the sudden quench (we attribute the fluctuations to the statistics of trajectories). For slower ramps, instead, the mean value of the order parameter tends exponentially towards the steady state, which approaches the free energy's global minimum of Eq.\ \eqref{freeenergy} for $\tau>10^4/\kappa$. We believe that this behaviour is determined by the ramp duration $\tau$ with respect to the time scale of the transient dynamics, and thus by the time the parameters $\alpha_n(t)$ spend close to the transition point. This conjecture is supported by the analysis of the time evolution of the single-particle kinetic energy shown in Fig.\ \ref{fig:ramp:p}, corresponding to the curves in Fig.\ \ref{fig:ramp:Theta}. For faster ramps it is similar to the sudden quench, exhibiting first a violent relaxation followed by a time interval where the dynamics is prevailingly coherent, and finally an exponential decay to the steady state value due to cavity cooling. In contrast, upon increasing the ramp duration towards slower ramps this transient regime disappears. Particularly, for the slowest ramp considered here, dissipation leads to quasi-adiabatic dynamics. Figure \ref{fig:ramp:St} shows the order parameters $\langle |\Theta_1(t)|\rangle$ and $\langle |\Theta_2(t)|\rangle$ at $t=3.77\times 10^6/\kappa$, where the curves of Fig.\ \ref{fig:ramp:Theta} have reached an asymptotic behaviour. Self-organization in the long-wavelength grating depends on the ramp duration $\tau$ and is found for $\tau>10^{4}/\kappa$. Note that short-wavelength order quantified by $\langle |\Theta_2(t)|\rangle$, in contrast, only slightly depends on the ramp duration.

On a microscopic scale, it seems that the reason for better long-wavelength ordering after slower ramps is that more time is spent close to the transition line ($\alpha_1=\alpha_2\sim1$), where the local minimum of the free energy is not deep enough to stably trap the system. In order to test this conjecture, we consider a two-step quench protocol which splits the sudden quench of Sec.\ \ref{subsec:quench_mixed} into two subsequent quenches: one at  $t=0$ from a paramagnetic to a ferromagnetic bistable phase, but close to the transition line: $\alpha_{1\rm int}=\alpha_{2\rm int}=1.1$. This quench shows a vanishing value of $P_t(\Theta_2<0)$ for sufficiently long times as in Fig.\ \ref{fig:mixed_quench:Steady}. The second sudden quench occurs after an elapsed time $\tau$ and goes from this intermediate point into $\alpha_{1f}=\alpha_{2f}=2$. The detuning $\Delta_c$ is kept constant during the evolution. 

\begin{figure}[h!]
\flushleft(a)\\
\center\includegraphics[width=0.4\textwidth]{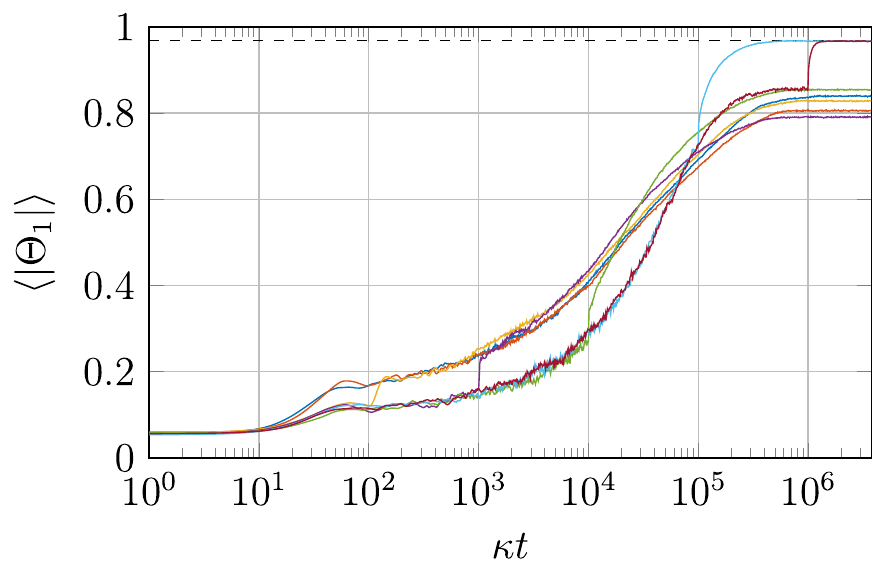}
\flushleft(b)\\
\center\includegraphics[width=0.4\textwidth]{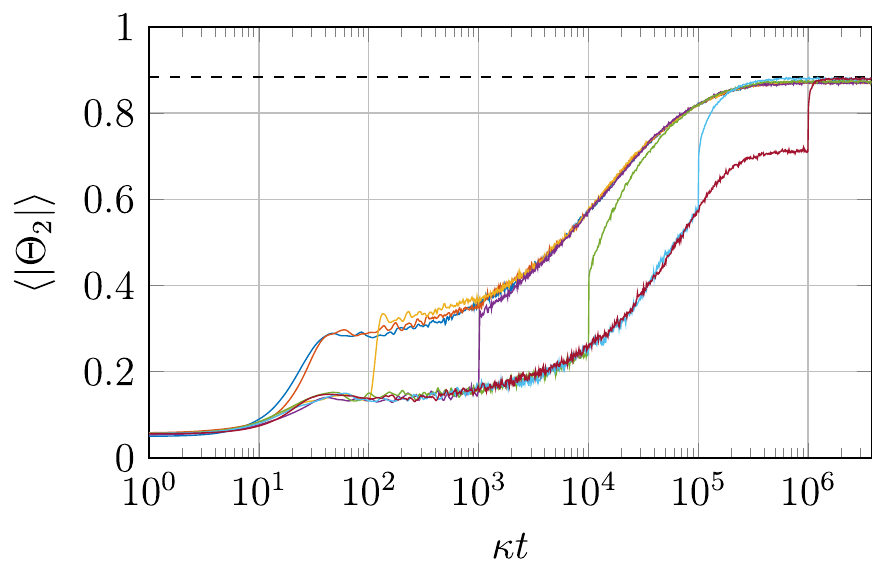}
\caption{
Dynamics of $\langle|\Theta_1|\rangle$ and $\langle|\Theta_2|\rangle$ as a function of time (in units of $1/\kappa$) for the two-step quench protocol. Here, the parameters $\alpha_n$ are suddenly ramped at $t=0$ from the initial values $\alpha_{1i},\alpha_{2i}\ll 1$  to $\alpha_{1\rm int}=\alpha_{2\rm int}=1.1$;  After a time interval $\tau$, there is a second quench from $\alpha_{1\rm int}=\alpha_{2\rm int}=1.1$ to $\alpha_{1f}=\alpha_{2f}=2$. The parameters are $\Delta_c=-\kappa$, $N=100$ with 250 trajectories, and  $\tau=1/\kappa$ (blue), $10/\kappa$ (red), $10^2/\kappa$ (yellow), $10^3/\kappa$ (purple), $10^4/\kappa$ (green), $10^5/\kappa$ (light blue) and $10^6/\kappa$ (dark red). The dashed horizontal lines show the value predicted by global minimum of the free energy, Eq. \eqref{freeenergy}, at $\alpha_1=\alpha_2=2$ and $\Delta_c=-\kappa$.
}
\label{fig:jump_ramp:Theta}
\end{figure}

\begin{figure}[h!]
\centering
\includegraphics[width=0.4\textwidth]{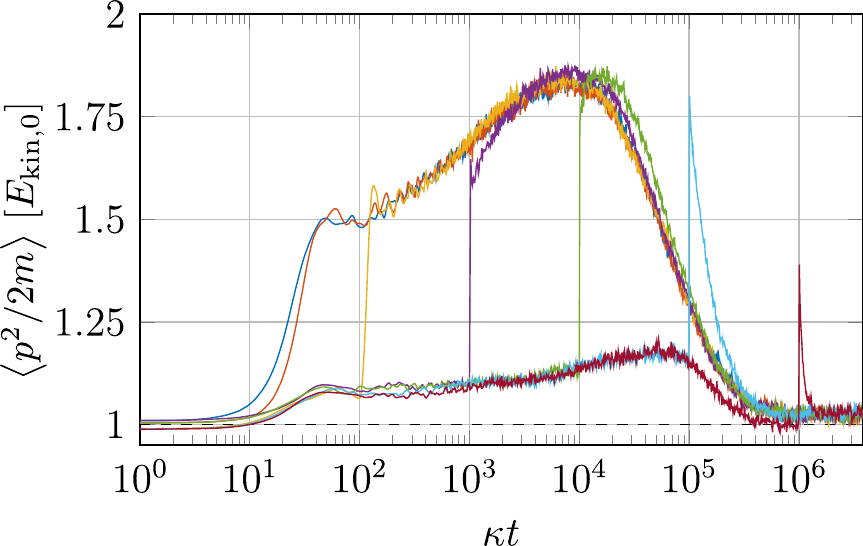}
\caption{Mean kinetic energy per particle, $\langle p^2/2m\rangle$ (in units of the asymptotic value $E_{\mathrm{kin},0}=\hbar \kappa/4$) as a function of time (in units of $1/\kappa$) for the same parameters and colour codes as in Fig. \ref{fig:jump_ramp:Theta}.
}
\label{fig:jump_ramp:p}
\end{figure}

Figure \ref{fig:jump_ramp:Theta} displays the time evolution of the mean absolute values of the order parameters for different time intervals $\tau$ elapsed between the two quenches. The order parameters undergo a first violent relaxation at $t=0$, when the first sudden quench occurs, and a second one immediately after the second quench (which looks like a jump in logarithmic scale). As expected, the larger the time elapsed between the two quenches, the closer the asymptotic value is to the one of the global minimum. Inspecting the dynamics of the kinetic energy in Fig.\ \ref{fig:jump_ramp:p} we observe that for large $\tau$ the atoms are cooled into the stationary state at $\alpha_n\sim1$. At this point of the phase diagram the free energy has two ferromagnetic global minima, while the nematic local minimum is very shallow. The system thus gets cooled close to the global minima of the free energy at $\alpha_n=2$, and remains trapped there after the second quench.

\begin{figure}[h!]
\centering
\includegraphics[width=0.4\textwidth]{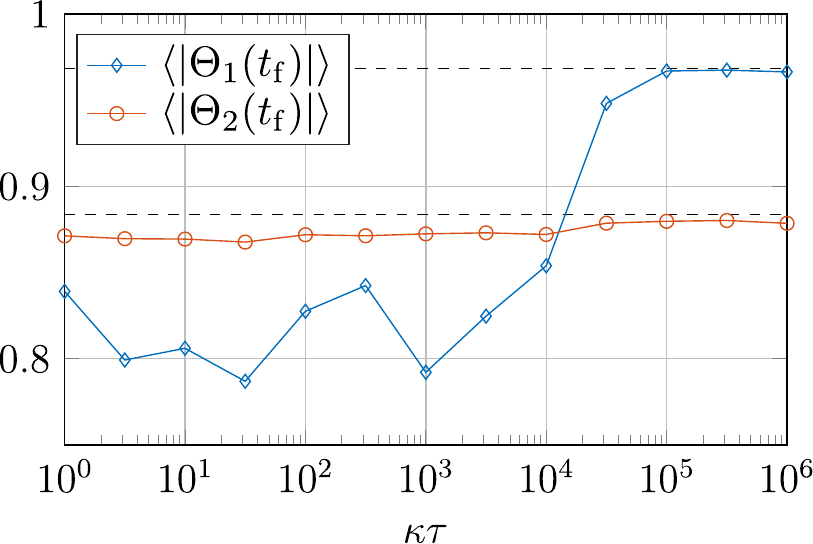}
\caption{The symbols correspond to $\langle |\Theta_1(t_{\mathrm{f}})|\rangle$ (blue) and $\langle |\Theta_2(t_{\mathrm{f}})|\rangle$ (red) at $t_{\mathrm{f}}=3.77\times 10^6/\kappa$ as a function of the time elapsed between the two quenches, $\tau$ (in units of $1/\kappa$), for the same parameters as in Fig. \ref{fig:jump_ramp:Theta}. The dashed horizontal lines show the steady state value predicted by the global minimum of the free energy, Eq. \eqref{freeenergy} at $\alpha_1=\alpha_2=2$ and $\Delta_c=-\kappa$.
}
\label{fig:jump_ramp:St}
\end{figure}

Figure \ref{fig:jump_ramp:St} displays the mean absolute value of the order parameters, as extracted from the numerical data at  $t=3.77\times 10^6/\kappa$, as a function of the time elapsed between the two quenches. Their value is compared to the predictions of the global minimum of the free energy at $\alpha_1=\alpha_2=2$ and $\Delta_c=-\kappa$. The behaviour is quite similar to the one observed when performing a linear ramp of corresponding duration, Fig. \ref{fig:ramp:St}. Dynamical ordering in the long-wavelength mode seems thus to require that the atoms are initially cooled close to the global minima. This is realised by means of the sufficiently large time $\tau$ spent close to the transition point. 

\section{Cooling into crystalline order}
\label{sec:mpemba}

We now analyse sudden quenches of the parameter $\alpha_n$ starting with different initial single-particle momentum widths. A possible realization is a quench in the detuning since $\Delta_c$ controls the steady state temperature, see Eq. \eqref{T}. By these means we consider quenches which could either lead to heating or cooling of the system to
the stationary temperature $T_0$, 
\begin{align}
k_BT_0=\frac{\hbar \kappa}{2},\label{T0}
\end{align} namely, the minimal temperature achieved by cavity cooling, corresponding to setting $\Delta_c=-\kappa$. Thereby we also consider initial thermal distributions which are spatially uniform and with temperature $T_{\mathrm{ini}}<T_0$. The initial momentum distribution we consider are Gaussian and their width is $\Delta p^2=m k_BT_{\text{ini}}$. 
\begin{figure}[h!]
\flushleft(a)\\
\center\includegraphics[width=0.4\textwidth]{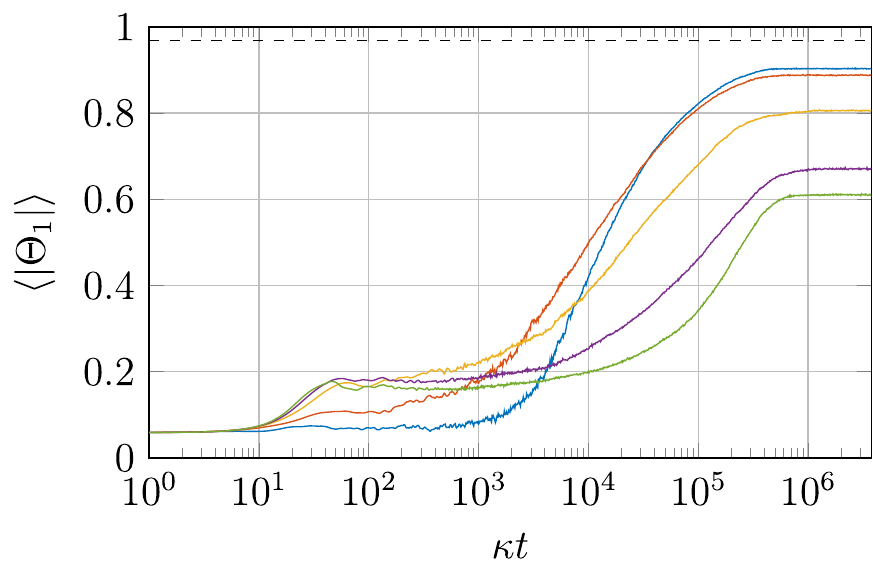}
\flushleft(b)\\
\center\includegraphics[width=0.4\textwidth]{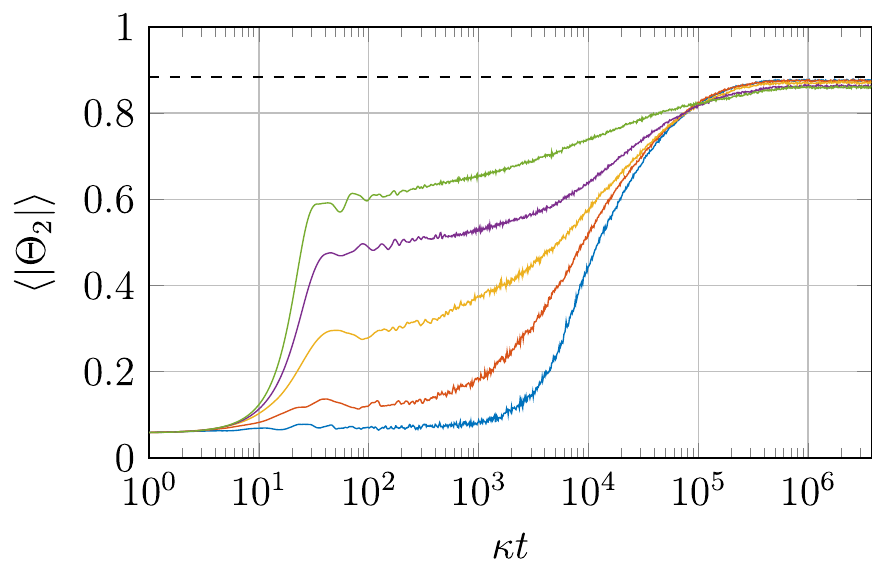}
\caption{Dynamics of (a) $\langle |\Theta_1|\rangle$ and (b) $\langle |\Theta_2|\rangle$ as a function of time (in units of $1/\kappa$) after a sudden quench at $t=0$ from $\alpha_{1i}=\alpha_{2i}=0$ and temperature $T_{\text{ini}}$ to $\alpha_{1f}=\alpha_{2f}=2$ and $\Delta_c=-\kappa$ (corresponding to the asymptotic temperature $T_0$), for $N=100$ and 250 trajectories. The different curves correspond to  $T_{\text{ini}}=5 T_0$ (blue), $2T_0$ (red), $T_0$ (yellow), $0.5 T_0$ (purple) and $0.1T_0$ (green). The dashed horizontal lines show the steady state value predicted by the global minimum of the free energy, Eq.\ \eqref{freeenergy}.
}
\label{fig:mpemba:Theta}
\end{figure}

\begin{figure}[h!]
\centering
\includegraphics[width=0.4\textwidth]{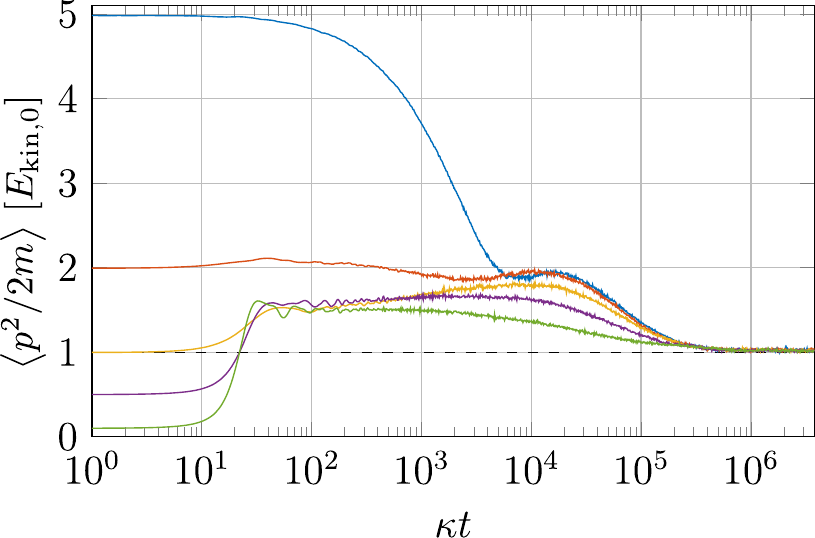}
\caption{
Mean kinetic energy per particle, $\langle p^2/2m\rangle$ (in units of $E_{\mathrm{kin},0}=\hbar \kappa/4$) as a function of time (in units of $1/\kappa$), for the same parameters and colour codes as in Fig. \ref{fig:mpemba:Theta}. The horizontal dashed line corresponds to the asymptotic value $\langle p^2/2m\rangle=E_{\mathrm{kin},0}$. 
}
\label{fig:mpemba:p}
\end{figure}

Figure \ref{fig:mpemba:Theta} shows the time evolution of the  mean absolute values of the order parameters for different values of $T_{\text{ini}}$ ranging from $0.1T_0$ up to $5T_0$. The asymptotic value of $\langle|\Theta_1|\rangle$ increases with the initial temperature: The hotter is initially the system, the smaller is the fraction of trajectories which remain trapped in the metastable, nematic state. The corresponding time evolution of mean kinetic energy per particle is displayed in Fig.\ \ref{fig:mpemba:p} and shows that for $T_{\text{ini}}=2T_0$ (and even more for $T_{\text{ini}}=5T_0$) the system stays relatively hot over time scales of the order of $10^4/\kappa$. For lower initial temperatures, instead, the system is heated by the energy released by the sudden quench before relaxation cools the atoms. 

As visible in Fig. \ref{fig:mpemba:Theta}, for samples initially cold a long-wavelength Bragg grating is formed faster than for hotter samples. In this case we recognize a three-stage dynamics like the one observed for the sudden quenches of the laser intensity, when a transient long-range order is established for times $t>10/\kappa$ and $t<10^3/\kappa$. For $t>10^3/\kappa$ dissipation becomes important and $\langle|\Theta_1|\rangle$ increases to a stationary value. This relaxation stage is also present for samples with initial temperatures larger than $T_0$, however, in this hotter case it is significant faster. Taking a threshold value $\left.\langle|\Theta_1|\rangle\right|_{\mathrm{thres}}=0.5$, we observe that buildup of long-wavelength order can take up to a hundred times shorter than for a cold initial state. This is reminiscent of the Mpemba effect in supercooled water \cite{Auerbach:1995,Brownridge:2011,Tao:2016,Jin:2015,Zhang:2013}. Its origin could be traced to a suppression of long-wavelength order if short-wavelength order is already established on a much faster time scale, visible in Fig.\ \ref{fig:mpemba:Theta}(b).

\begin{figure}[h!]
\centering
\includegraphics[width=0.37\textwidth]{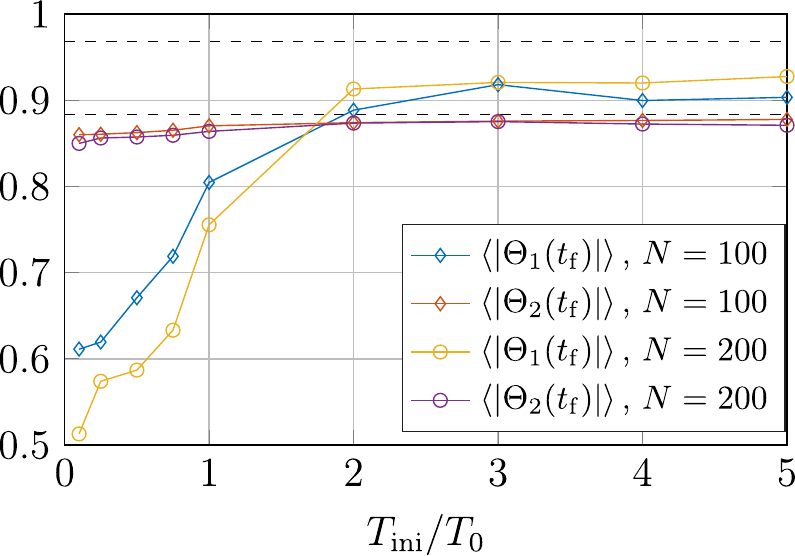}
\caption{The symbols correspond to the values of  $\langle |\Theta_1(t_{\mathrm{f}})|\rangle$ (blue) and $\langle |\Theta_2(t_{\mathrm{f}})|\rangle$ (red) at $t_{\mathrm{f}}=3.77\times 10^6/\kappa$ as a function of $T_{\text{ini}}$ (in units of $T_0$), for the same parameters as in Fig. \ref{fig:jump_ramp:Theta}, but for $N=200$ particles (125 trajectories). See box for the color code. The dashed horizontal lines show the steady state value predicted by the global minimum of the free energy, Eq. \eqref{freeenergy}.
}
\label{fig:mpemba:St}
\end{figure}

In Fig.\ \ref{fig:mpemba:Theta} (a) we observe that the final value of $\langle|\Theta_1|\rangle$ does not coincide with its predicted stationary value even after very long cooling times. This can also be seen in Fig.\ \ref{fig:mpemba:St}, which shows the mean absolute value of the order parameters at $t=3.77\times10^6/\kappa$ as a function of the initial temperature for $N=100,200$. One would expect that $\langle|\Theta_1|\rangle$ should have reached a constant value corresponding to the stationary state. Apparently this is not the case and even for finite $N$ a significant fraction of trajectories converges to and remains in the local minimum. This behavior gets much less pronounced, if the initial temperature lies above a certain threshold set by the energy released by the quench itself.

\section{Comparison between different numerical approaches}
\label{sec:comparison}

The discussion of this paper has been based on results obtained by numerical integration of the stochastic differential equations \eqref{com} and on their comparison with the corresponding analytical model. Both rely on the validity of the so-called bad cavity limit, where cavity damping is the fastest time scale, and in particular on treating retardation as a small parameter in the dynamics. This regime allows one to systematically describe the quantum fluctuations of the cavity degrees of freedom by eliminating the cavity variables from the equations of motion of the external degrees of freedom. We now compare these predictions with the ones of stochastic differential equations derived in Ref.\ \cite{Domokos:2001} where the cavity degrees of freedom are treated in the semi-classical limit but included at all orders of the retardation expansion. These stochastic differential equations are here extended to our setup composed of two cavity modes \cite{Torggler:2014} and read:
\begin{subequations}
	\begin{align}
	&dx_j=\frac{p_j}{m}dt\,,\label{xwcav}\\
	&dp_j=\sum_{n=1,2}2\hbar nk S_n \mathcal{E}_{n,r}\sin(nkx_j)dt\,,\label{pwcav}\\
	&d\mathcal{E}_{n,r}=(-\Delta_n\mathcal{E}_{n,i}-\kappa_n \mathcal{E}_{n,r})dt+d\xi_{n,r}\,,\label{Er}\\
	&d\mathcal{E}_{n,i}=(\Delta_n\mathcal{E}_{n,r}-\kappa_n \mathcal{E}_{n,i}-NS_n \Theta_n)dt+d\xi_{n,i}\,,\label{Ei}
	\end{align}
	\label{com+cavity}
\end{subequations}
\noindent where $\mathcal{E}_{n,r}={\rm Re}\{\mathcal{E}_{n}\}$ and  $\mathcal{E}_{n,i}={\rm Im}\{\mathcal{E}_{n}\}$ are the real and imaginary part of the positive-frequency component of the cavity field mode $n=1,2$. The Wiener processes $d\xi_{n,i},d\xi_{n,r}$ have vanishing first moment, $\langle d\xi_{n,i}\rangle=0=\langle d\xi_{n,r}\rangle$, while the second moments fulfill $\langle d\xi_{n,i}d\xi_{m,i}\rangle=\delta_{nm}\kappa/2dt$, $\langle d\xi_{n,r}d\xi_{m,r}\rangle=\delta_{nm}\kappa/2dt$, and $\langle d\xi_{n,r}d\xi_{m,i}\rangle=0$. 

The results of the simulations based on the two approaches for a single-mode cavity show good agreement. For the two-mode cavity we generally find qualitative agreement. Quantitative discrepancies are found in general for the momentum distribution: The simulations based on Eq.\ \eqref{com+cavity} predict for certain parameters samples whose temperature is 10\% hotter than the one obtained with Eq.\ \eqref{com}. Small differences are found also for the order parameters after the quenches into the bistable phase. 

\begin{figure}[h!]
	\flushleft (a)\\
	\center\includegraphics[width=0.4\textwidth]{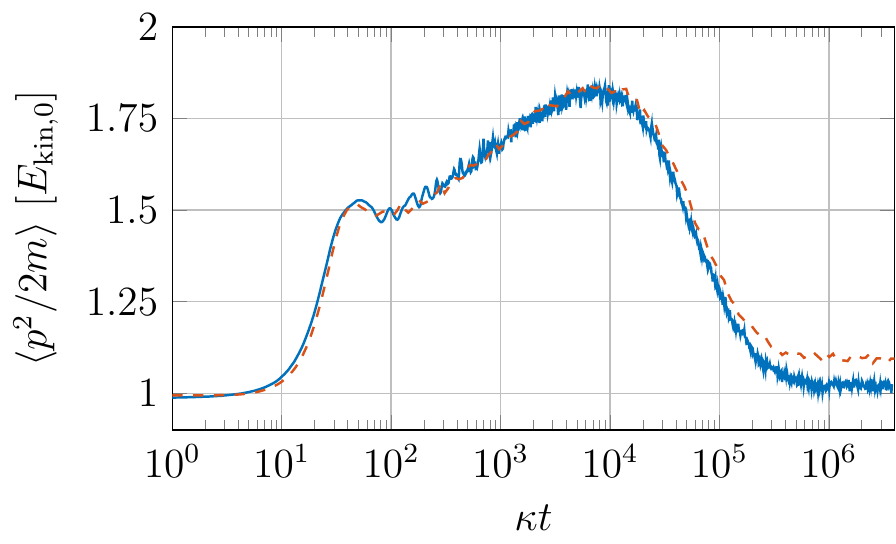}
	\flushleft (b)\\
	\center\includegraphics[width=0.4\textwidth]{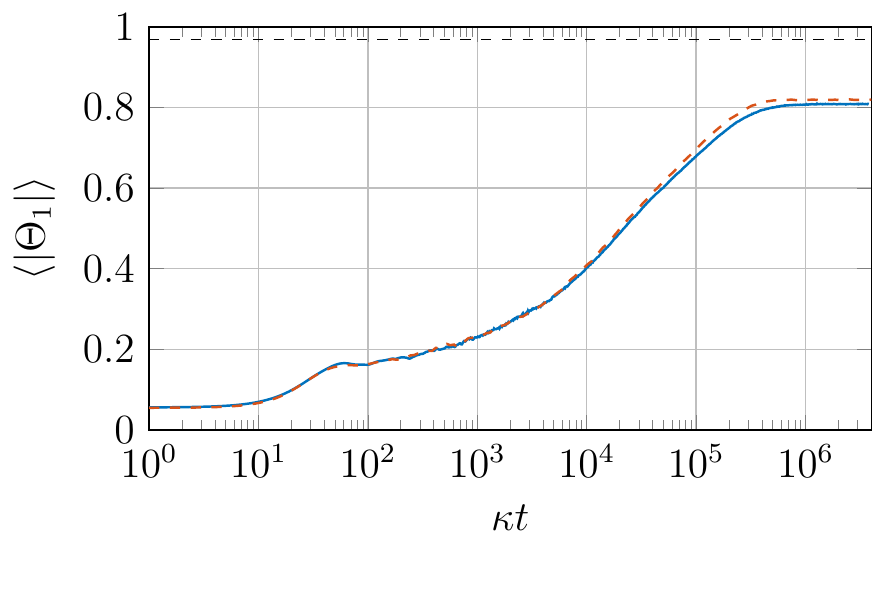}
	\flushleft (c)\\
	\center\includegraphics[width=0.4\textwidth]{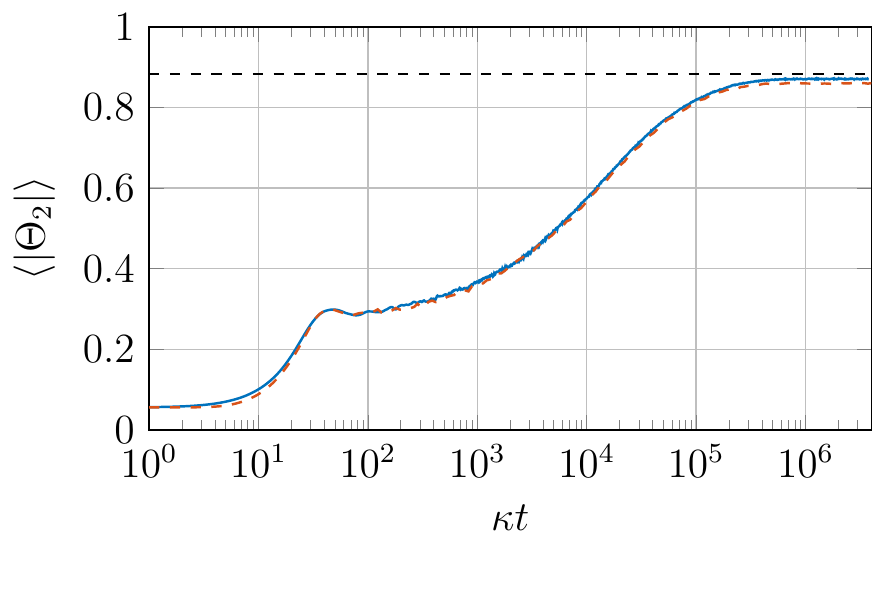}
	\caption{Dynamics of (a) single-particle kinetic energy  (in units of $E_{\rm kin,0}$), (b) $\langle|\Theta_1|\rangle$, and (c) $\langle|\Theta_2|\rangle$  as a function of time (in units of $1/\kappa$) following a quench at $t=0$ from $\alpha_{1i}=\alpha_{2i}\ll 1$ to $\alpha_1=\alpha_2=2$, for $\Delta_c=-\kappa$ and $N=100$. The blue (red) lines correspond to the simulations using Eq.\ \eqref{com}(Eq.\ \eqref{com+cavity}). The black dashed lines mark the values of the order parameters obtained by the free energy, Eq.\ \eqref{freeenergy}. In (a) the blue (red) line corresponds to 250 (500) trajectories. In (b) and (c) the blue and red lines correspond to 1000 (500) trajectories. Note, that a quench from $\alpha_{1},\alpha_{2}$ as performed in Sec.\ \ref{subsec:quench_mixed} for fixed $\Delta_c=-\kappa$ corresponds for the simulation of Eq. \eqref{com+cavity} to a quench in the pumping strengths $S_{n}$ such that $NS_{n}^2=\alpha_n\kappa^2$. At time $t_{\mathrm{f}}=3.77\times 10^6/\kappa$ we observe that 16.9\% (15.4\%) of the trajectories are in nematic phase using Eq.\ \eqref{com} (Eq.\ \eqref{com+cavity}).
		\label{Fig:comparison}}
\end{figure}

Figure \ref{Fig:comparison} shows a representative result of the discrepancies found after the quench protocol discussed in Sec. \ref{subsec:quench_mixed}.
The two simulations predict different stationary values for both kinetic energy and the order parameters. We believe that this discrepancy is due to retardation effects, which are neglected in the approach of Eq. \eqref{com} and become relevant when the atoms are trapped at tight minima. 

In order to test our conjecture we use the prediction of the kinetic theory of Refs. \cite{Niedenzu:2011,Griesser:2012}, where the temperature of the stationary thermal distribution was corrected by the contribution due to the atoms' localization at the minima of the self-organized lattice,
\begin{align}
k_{B}\tilde{T}=\hbar \frac{\Delta_c^2+\kappa^2}{4|\Delta_c|}+\hbar\frac{\omega_0^2}{|\Delta_c|}\,.\label{temperature}
\end{align}
Here, $\omega_0$ is the frequency of oscillation about the lattice minima in the harmonic approximation. It can be estimated using Eq.\ \eqref{pwcav} and imposing the equality
\begin{align*}
dp_j\approx\sum_{n=1,2}2\hbar (nk)^2 S_n \mathcal{E}_{n,r}x_jdt \equiv - m\omega_0^2x_jdt\,.
\end{align*}
This delivers an analytic estimate of the frequency
\begin{align*}
\omega_0^2=\omega_r\frac{\Delta_c^2+\kappa^2}{-\Delta_c}\left(\alpha_1\Theta_1+4\alpha_2\Theta_2\right),
\end{align*}
where we used Eq. \eqref{eq:nthreshold}. For the parameters of the quench in Fig. \ref{Fig:comparison}, with $\Delta_c=-\kappa$ and $\alpha_1=\alpha_2=2$, we obtain
$k_B\tilde{T}\approx 1.1k_BT_0$, where $T_0$ is the temperature given in Eq. \eqref{T0}.
Indeed, this corrected value of the final temperature is in good agreement with the discrepancy observed in Fig. \ref{Fig:comparison} (a).

This hypothesis is also consistent with the discrepancy observed in the asymptotic values of the order parameters.  In fact, stationary temperature and the final values of the order parameters are related: the stationary values of the order parameters are determined by the parameters $\alpha_1,\alpha_2$ (cif.\ \cite{Keller:2017}) and thus depend on both field intensities as well as detunings, see Eq.\ \eqref{eq:nthreshold}. According to this hypothesis, the asymptotic values of the order parameters for the simulation using Eq.\ \eqref{com+cavity} should be the ones corresponding to the system's parameters with the corrected temperature $\tilde{T}$, hence we shall minimize the free energy of Eq.\ \eqref{freeenergy} using $\tilde{\beta}=1/(k_B\tilde{T})$, Eq. \eqref{temperature}, instead of $1/(k_BT_0)$. This is equivalent to rescale the phase diagram in Fig.\ \ref{fig:phase_diagram}(a) using the prescription $\tilde{\alpha}_n=\alpha_nT_0/\tilde{T}<\alpha_n$, and results in a smaller stationary value of the order parameter which is consistent with the discrepancies visible in Fig.\ \ref{Fig:comparison} (b) and (c).

\section{Conclusions}
\label{sec:conclusions}

In this work we have studied the semi-classical dynamics of atoms interacting with two cavity modes after quenches of the intensity and/or frequency of the pumping lasers. In the quench protocols the laser parameters were varied across transition lines separating a disordered from an ordered self-organized phase. We could verify numerically that the states reached at the asymptotics of the dynamics correspond to the minima of the free energy of a corresponding thermodynamic description developed in Ref.\ \cite{Keller:2017}. This picture is further confirmed by the comparison with numerical simulations based on different initial assumptions. This analysis shows, in particular, that trapping of the system in local minima of the free energy crucially depends on the initial temperature and on the cooling rate.

We observe, in addition, that the system can be trapped in metastable configurations for transient times which cannot be understood in terms of the effective thermodynamic description. For hundreds of particles the lifetime of these states is about four orders of magnitude longer than the cavity lifetime, and is expected to increase with $N$. They share analogies with metastable configurations found in the GHMF when performing quenches in the microcanonical ensemble \cite{Teles:2012}. Since the phase diagrams of GHMF and the model here considered can be formally mapped into one another \cite{Keller:2017}, we conjecture that these metastable configurations could be due to the coherent dynamics. This conjecture can be tested by means of a mean-field analysis as the one performed in Ref.\ \cite{Jaeger:2016} for a single mode cavity.

Interestingly, when the initial temperature of the atomic ensemble is different from the stationary temperature of cavity cooling, we observe that the final magnitude of asymptotic order changes. In particular when the initial temperature is even lower than the predicted cavity cooling temperature, the probability that the systems remains trapped in metastable configurations is further increased. This reminds of the behavior of supercooled water \cite{Auerbach:1995,Brownridge:2011,Tao:2016,Jin:2015,Zhang:2013}.

Here we have considered the very special case of two commensurate modes. While this already highlights many generic properties of the dynamics, future considerations certainly should include the case in which the wavelength of the cavity modes are incommensurate \cite{Habibian:2013}, so that the ordering mechanisms are much more strongly competing and a multitude of meta-stable states can form. A further interesting direction is operation with much colder temperatures or in the side-band resolved cooling regime \cite{Kramer:2014}. Here it is intriguing to consider in which form meta-stable states survive deep in the quantum regime. Besides diffusion they could be depleted via tunneling and atom-field entanglement plays an important role in this dynamics   \cite{Maschler:2007}, a process which should also be relevant in closely related schemes of simulated quantum annealing \cite{Torggler:2017}.

\acknowledgments

The authors thank Tobias Donner, Sebastian Kr\"amer, and Francesco Rosati for stimulating and helpful discussions. This work was supported by the German Research Foundation (DFG, DACH project "Quantum crystals of matter and light") and by the European Commission (ITN network "ColOpt"). V. T. and H. R. are supported by Austrian Science Fund Project No. I1697-N27. T. K. and V. T. contributed equally to this work.

\bibliography{bibliography}

\end{document}